\documentclass[%
reprint,
amsmath,amssymb,
aps,
]{revtex4-2}

\usepackage{graphicx}
\usepackage{dcolumn}
\usepackage{bm}
\usepackage{float} 
\usepackage{url} 
\usepackage[colorlinks=true, citecolor=blue, linkcolor=blue, urlcolor=blue]{hyperref}
\usepackage{qcircuit}
\usepackage{amssymb}
\usepackage{subcaption}
\usepackage{svg}
\usepackage{appendix} 
\usepackage{cleveref}
\usepackage[figuresleft]{rotating}
\usepackage{float}
\usepackage{braket}
\usepackage{adjustbox}
\usepackage{caption}  
\usepackage{ragged2e}
\usepackage{xcolor}

\usepackage{cellspace}
\definecolor{JLHblue}{RGB}{25,25,125}
\definecolor{ZJWorange}{RGB}{237, 140, 12}
\definecolor{ODred}{RGB}{166, 18, 10}

\begin{document}

\preprint{APS/123-QED}

\title{Quantum Sinusoidal Neural Networks}

\author{Zujin Wen}
\affiliation{Department of Physics, City University of Hong Kong, Tat Chee Avenue, Kowloon, Hong Kong SAR, China}

\author{Jin-Long Huang}
\affiliation{Department of Physics, City University of Hong Kong, Tat Chee Avenue, Kowloon, Hong Kong SAR, China}

\author{Oscar Dahlsten}
\email{oscar.dahlsten@cityu.edu.hk}
\affiliation{Department of Physics, City University of Hong Kong, Tat Chee Avenue, Kowloon, Hong Kong SAR, China}
\affiliation{Shenzhen Institute for Quantum Science and Engineering, SUSTech, Nanshan District, Shenzhen 518055, China}
\affiliation{Institute of Nanoscience and Applications, Southern University of Science and Technology, Shenzhen 518055, China}

\date{\today}

\begin{abstract}
We design a quantum version of neural networks with sinusoidal activation functions and compare its performance to the classical case. We create a general quantum sine circuit implementing a discretised sinusoidal activation function. Along the way, we define a classical discrete sinusoidal neural network. We build a quantum optimization algorithm around the quantum sine circuit, combining quantum search and phase estimation. This algorithm is guaranteed to find the weights with global minimum loss on the training data. We give a computational complexity analysis and demonstrate the algorithm in an example. We compare the performance with that of the standard gradient descent training method for classical sinusoidal neural networks. We show that (i) the standard classical training method typically leads to bad local minima in terms of mean squared error on test data and (ii) the weights that perform best on the training data generalise well to the test data. Points (i) and (ii) motivate using the quantum training algorithm, which is guaranteed to find the best weights on the training data. 
\end{abstract}

\maketitle

\section{\label{sec:1}Introduction}
Sinusoidal activation functions of artificial neural networks have a large degree of freedom via their non-trivial second and higher derivatives. Such sinusoidal neural networks have been shown to be suitable for modelling natural signals and have demonstrated their effectiveness across a wide range of applications, including image fitting, video, and audio representation; 3D shape reconstruction via signed distance functions; solving partial differential equations; image editing and inpainting; and generalising across a family of signals via learning shared structure (i.e., learning priors over function spaces using hypernetworks)~\cite{SIREN, FAIL, UNDERSTAND}. 

Sinusoidal neural nets are commonly trained via gradient descent-based methods~\cite{SinNNTaming, SIREN}.  Bad local minima pose a challenge to gradient descent training of sinusoidal neural networks~\cite{SinNNTaming, SIREN}, as with  neural networks with commonly used activation functions like sigmoid and ReLU~\cite{LOCAL}


Training in quantum superposition via quantum search provides a method for both avoiding local minima as well as a speed advantage over brute-force search. Quantum search~\cite{GRO} has been applied to reduce the query complexity of classification~\cite{gbls} and for faster reinforcement learning~\cite{quantum-q-learning}. Quantum search has been adapted for optimising classical neural nets~\cite{QBNN} and for quantum neural nets~\cite{wan2017quantum,qcnn, training-qnn, power-of-qnn, overparametrization-qnn} in Ref.~\cite{quantumTrainingLiao2024} as a method to avoid the barren plateau~\cite{barrenPlateaus} problem of quantum neural nets and to outperform brute force in terms of speed. These promising results suggest that quantum training may circumvent the challenge of bad local minima in sinusoidal neural nets and thereby outperform classical sinusoidal neural nets.


We here tackle this question by designing quantum sinusoidal neural nets which can be trained in a superposition of weight states. We introduce a {\em discrete} SinNN model (DSinNN) with discrete weights and outputs, amenable to qubit implementation. We design a quantum sine circuit which implements an associated discretised sinusoidal function. This circuit can accept superposition of inputs. In the quantum sinusoidal neural net the weights are also (sets of) qubits and may be in a superposition of states. The amplitudes of the weights that perform well on the training data are gradually amplified during the training. The output of the neuron for a given input is compared to the desired output and a phase $e^{i\Pi/N}$ is added if they match, where $N$ is the number of training samples. 
If the output is always correct for given weights the final state after all the training samples are fed through is $e^{i\Pi N/N}=-1$ and, more generally, phase estimation~\cite{PE} is used to decide whether a given phase should count as -1 or 1. The resulting sub-circuit acts as an oracle in a quantum search, assigning -1 to those strings that should be amplified. Via results concerning quantum search,  the algorithm is then analytically guaranteed to obtain a global minimum with high probability~\cite{TIGHTBOUND}. We term the overall algorithm, which is defined for any finite-sized scenario, QSinNN.

We demonstrate QSinNN on a tractable toy model DSinNN model with two hidden neurons via classical simulation of 20 logical qubits. The simulation shows explicitly how the quantum algorithm finds the global minimum.


To evaluate the need for QSinNN, we simulate large classical SinNN and DSinNN training via (stochastic) gradient descent. We find that for DSinNN in 85\% of training instances (random seeds), across many architectures, gradient descent leads to training results whose mean squared error on the test data is more than twice that of the minimal (encountered) mean squared error. For the continuous case, the corresponding number is 95\%. Thus, QSinNN implemented on large neural nets will significantly outperform gradient descent training for the tasks considered here. 

The classical DSinNN network defined along the way may have independent value.
Discrete neural nets are significantly more hardware efficient than continuous neural nets~\cite{BNN}.  When weights and activations of DSinNN are represented in higher precision, the model converges to SinNN. 

The results point to the potential for marrying bulky, expensive quantum cloud computers with small classical edge computing devices. The idea is that Quantum cloud operators, such as IBM Quantum, Amazon Braket, Microsoft Azure Quantum and Origin Quantum~\cite{IBM, originqc}, would optimise the design of the mass-produced classical DSinNN hardware, which then executes the online inference.

\section{Classical sinusoidal neural networks and their training}
In this section, we first describe classical sinusoidal neural networks and how we discretise them. We then demonstrate the bad local minima issues with gradient descent training methods of different size of the networks.

\subsection{Sinusoidal activation functions}
A key characteristic of a neural net is the activation function. Each layer of a neural network consists of two consecutive operations: linear and nonlinear transformations, with the nonlinear transformations termed activation functions. Various functions have been used for the nonlinear transformation, such as Sigmoid~\cite{back-prop}, ReLU~\cite{ReLU}, ELU~\cite{ELU}, Leaky ReLU~\cite{LeakyReLU}, GELU~\cite{GELU}, and SiLU~\cite{SiLU}. There are also extensive studies using periodic functions as activation. Fourier neural networks incorporate the Fourier transformation into neural networks~\cite{fourierNN, fourierNNSurvey}. Periodic nonlinearities have been used to represent images and sequential data~\cite{SinNNHandwritten, SinNNTaming, SinRNNSequential, SinRNNPower, SinRNNStability}. Recently, periodic activation functions like \textit{sine} functions have been shown to accurately represent complex natural signals, with wide applications in processing images, sound, and video, and solving differential equations~\cite{SIREN}. The representational power of such activation functions originates from their capability to model higher-order derivatives of spatial and temporal signals~\cite{SIREN}.

A sinusoidal Neural Network (SinNN) uses the sine function as an activation function~\cite{SIREN}. The output $z'_j$ of a layer of neurons is obtained from inputs $z_i$ by
\begin{equation}
    z'_j =  \sin \sum_i  \left(w_{i j} \cdot z_i \right). 
\end{equation}
In general, the output $\hat{y}$ can be a non-linear function of the previous layer's outputs, like softmax, but here, for simplicity, we simply add those outputs.

\subsection{Classical discrete sinusoidal neural networks}

Binary Neural Networks (BNNs)~\cite{BNN} are a key class of discrete neural networks where the weights and activated values of neurons are constrained to binary values (e.g.\ -1 and 1). BNNs are emerging as a noteworthy development due to their hardware efficiency and effectiveness~\cite{BNN, BNNSurvey, BNN2}. BNNs significantly reduce computational demands and memory usage, making them particularly suitable for deployment in edge devices and resource-constrained environments. BNNs can be seen as a modification of traditional neural networks, designed to maximise efficiency with minimal impact on performance. BNNs are a special case of discretised, also called quantised, neural networks~\cite{quantizationSurveyBook, quantizationSurveyArxiv}, wherein more than one bit is used to represent an individual weight more generally.

We adopt a similar discretisation procedure as in~\cite{BNN} for the SinNN model. The new model is called DSinNN. For each hidden layer except the last in DSinNN, inputs $z_i$ are, in the forward pass, transformed into output $z'_j$ according to 
\begin{equation}\label{eq:DSinNN_forward}
    z'_j = \text{D} \left( \sin \sum_i  \left[ \lambda\, \text{D}(w_{ij}) \cdot z_i \right] \right),
\end{equation}
where 
\begin{equation}
\text{D}(x) \equiv  
\begin{cases} 
-1 & \text{if } x < 0 \\
1 & \text{if } x \ge 0 ,
\end{cases}
\label{eq:D}
\end{equation}

$w_{i j}$ is real-valued, and $\lambda$ is set to 1 for the first hidden layer and a tunable scaling constant for deeper hidden layers. For the last layer, the output is simply
\begin{equation}\label{eq:DSinNN_last}
    \hat{y}= \sum z_j',
\end{equation}

The Mean Square Error is used as the loss function to update weights. However, since the gradient of the sign function is zero almost everywhere, back-propagated errors will also be zero almost everywhere. To avoid this obstacle, the Straight-Through Estimator (STE) was proposed in~\cite{BNN} to let back-propagation bypass the discretisation procedure, such that real-valued weights could be properly updated in the training process. In this paper, we choose an identity function $f(x)=x$ for STE for simplicity, instead of the hardtanh used in~\cite{BNN}.

\subsection{Bad local minima in a small SinNN}
We develop a simple illustrative example to demonstrate the presence of bad local minima in SinNN. This model only has one scalar input and one scalar output with one hidden layer of two neurons. For each input $x$, model output is $\hat{y} = \sin (w_1 x) + \sin (w_2 x)$. 

This model is trained with four input-output pairs in a single batch. Thus the MSE loss averaged over this batch is $\frac{1}{4}\sum^{4}_{1}(y_i-\hat{y_i})^2$, where the index $i$ is over samples in the batch. Four input-output pairs $(x_i, y_i)$ are generated by the target function $y = \sin(1 \cdot x) + \sin (1 \cdot x)$:
\begin{center}
\begin{tabular}{c c c c}
    $(-\frac{3 \pi}{2}, 2),$ & $(-\frac{\pi}{2},-2),$ & $(\frac{\pi}{2},2),$ & $(\frac{3 \pi}{2}, -2)$.
\end{tabular}
\end{center}
Thus, the optimal weights are $(w_1, w_2) = (1, 1)$.

\begin{figure}[htbp!]
    \centering
    \begin{subfigure}[b]{1.0\linewidth}
    \includegraphics[width=\linewidth,page=1]{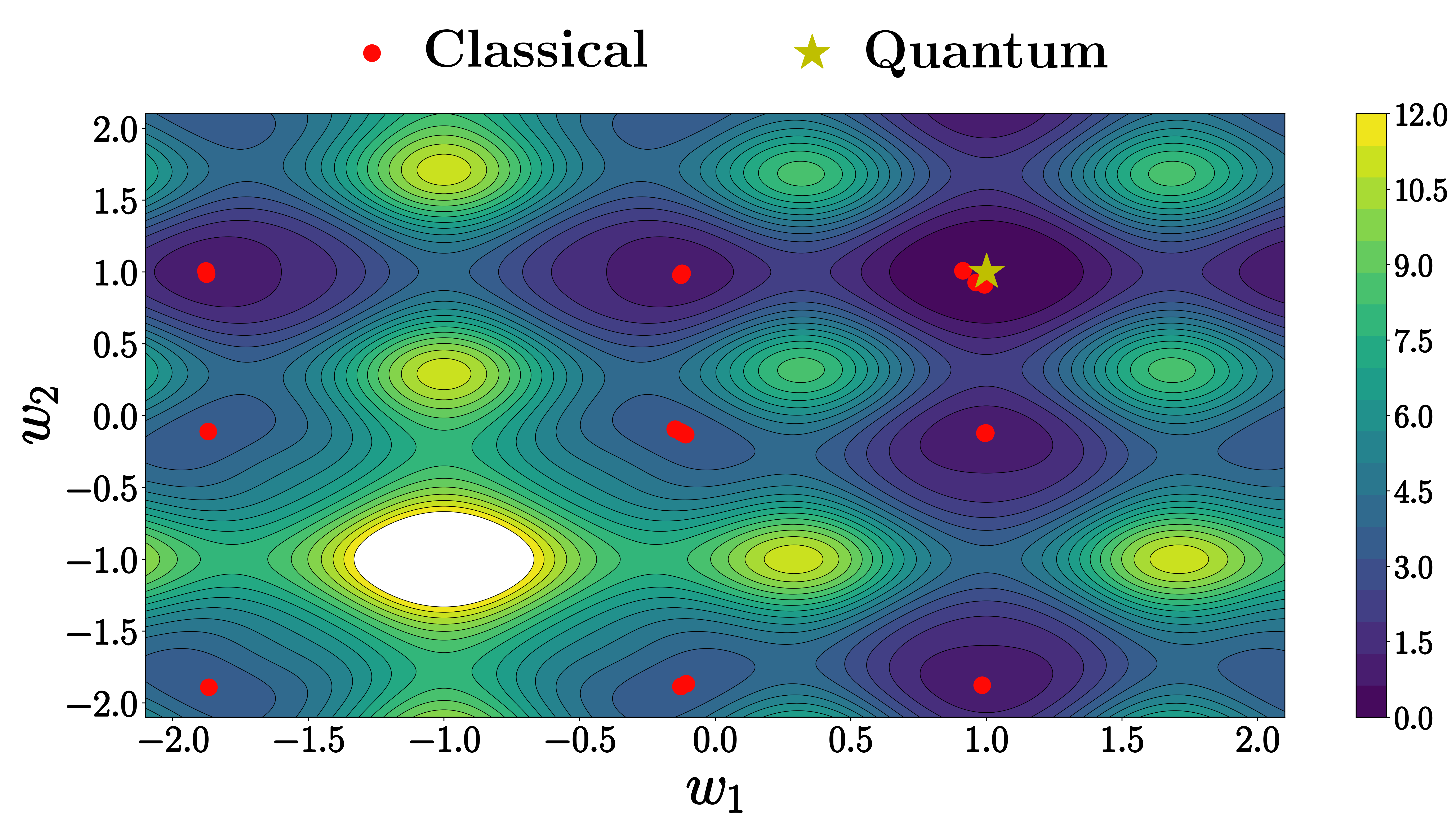}
    \caption{}
    \label{fig:bad_local_minima_SinNN}
    \end{subfigure}
    \hfill
    \begin{subfigure}[b]{1.0\linewidth}
    \includegraphics[width=\linewidth,page=1]{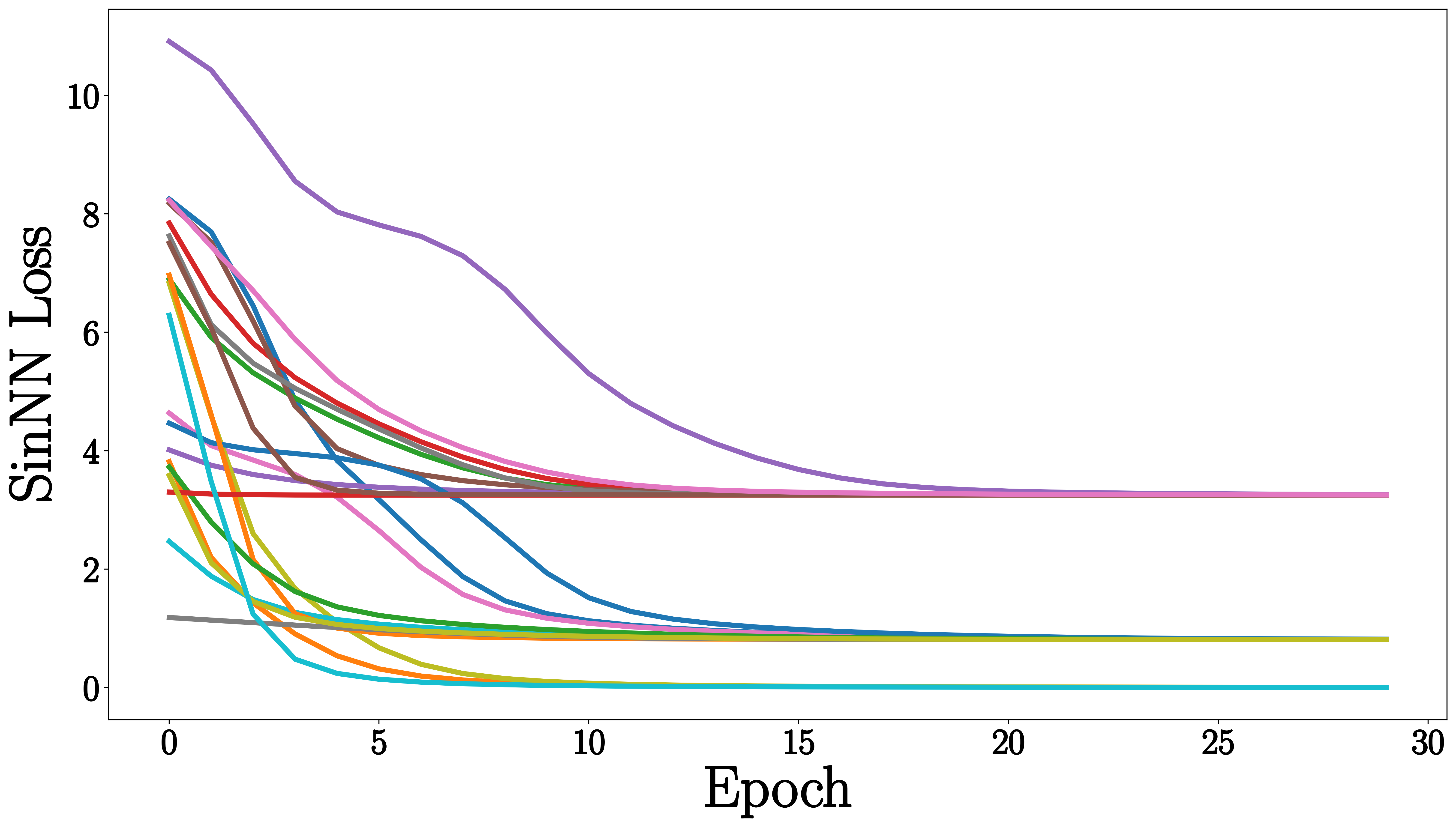}
    \caption{}
    \label{fig:SinNN_Loss_Epoch}
    \end{subfigure}
    \captionsetup{justification=raggedright,singlelinecheck=false}
    \caption{
    \textbf{Training of a simple SinNN model.}
        (a) Loss landscape for SinNN. The white regions' losses are greater than 12. There is 1 global minimum and 8 local minima in the figure.
        (b) Loss history for classical training of SinNN as a function of epochs for 20 different initial weights. Most of the lines do not reach the global minimum.
        }
\end{figure}

The contour plot of the loss landscape is presented in \autoref{fig:bad_local_minima_SinNN}. White regions are mountains with a loss greater than 12, and darker regions are valleys corresponding to a smaller loss. From the plot, we can observe 8 shallow valleys where the surrounding areas are higher, clearly indicating that these valleys are local minima, and only one global minimum exists, at location $(1, 1)$. Moreover, since the sine function is periodic, local minima will arise periodically, meaning that they can be found throughout the landscape. 

In \autoref{fig:SinNN_Loss_Epoch}, we plot the loss as a function of the number of epochs for 20 different initial weights. In most cases, the weights get stuck at local minima with a higher loss than the global minimum. In this situation, gradient descent has great difficulty in training neural networks to reach global minima.

\subsection{Bad local minima in large SinNNs}
As noted in~\cite{SIREN}, the performance of sinusoidal neural networks (SinNNs) is highly sensitive to weight initialisation, and insufficient care in this process can easily result in poor training stability and slow convergence. We have similarly observed, but do not elaborate here, that initialising the weights of a multi-layer SinNN over a large interval often causes the training loss to oscillate strongly across epochs and fail to converge. Given the widespread adoption of the initialisation scheme proposed in~\cite{SIREN}, we take it as a standard baseline in our experiments to systematically and numerically investigate the severity of the bad local minima problem. 

Specifically, the initialisation scheme is as follows. Assume that the input \(x\) is uniformly distributed over the interval \([-1, 1]\). For all hidden layers except the first, the weights are initialized from a uniform distribution $w_j \sim \mathcal{U}\left(-\sqrt{\frac{6}{n}}, \sqrt{\frac{6}{n}}\right)$, where \(n\) is the number of input connections (fan-in) to the neuron. For the first layer, the weights are initialized as
$w_i \sim \mathcal{U}\left(-1, 1\right)$, and then scaled by a factor $\lambda_0 = 20$, i.e., the first layer computes $\sin(\lambda_0 \cdot w_i x)$. This high-frequency scaling causes the sine activation to traverse multiple periods over the input domain, enabling the network to explore a broader range of high-frequency representations from the very beginning of training. This improves its capacity to model fine detail and contributes to stable training dynamics even in deep sinusoidal architectures.

To further illustrate the operation of neural networks with sinusoidal activation functions, we consider a more complex example than that in the previous subsection: fitting the target function $y=\sin{x}+\sin{2x}+\sin{3x}+\sin{4x}+\sin{5x}$. In our model, we adopt a ``decreasing" architecture, where the number of neurons in each hidden layer progressively decreases after the first layer. Specifically, we first define the number of neurons in the first hidden layer and the total number of hidden layers. Then, in each subsequent layer, the number of neurons is reduced by one in each subsequent layer until it reaches five, after which it remains fixed at five for any remaining hidden layers. This structure guarantees that the network's output covers the interval $[-5, 5]$, consistent with the range of $y\in[-5, 5]$.

We uniformly sample 200 input-output pairs $(x, y)$ from the interval $x\in [-\frac{7\pi}{2}, \frac{7\pi}{2}]$. Among these, 80\% are randomly selected as training data and the remaining 20\% are used for testing. Prior to training, all input values are normalised to the range $[-1, 1]$ to ensure consistency with the input domain assumed by the initialisation scheme.

We train our model with an initial weight scaling factor $\lambda_0 = 20$, 200 random seeds, a learning rate of 0.002, and 300 epochs. A local minimum is considered ``bad" if its final loss is significantly larger than the global minimum—more precisely, if it exceeds twice the value of the global minimum. To further reduce the risk of misclassification in cases where the global minimum is close to zero, we apply an additional condition that the final loss must also exceed the global minimum by at least 0.0001 to be classified as a bad local minimum.

\begin{figure}[!htbp]
    \centering
    \includegraphics[width=0.95\linewidth]{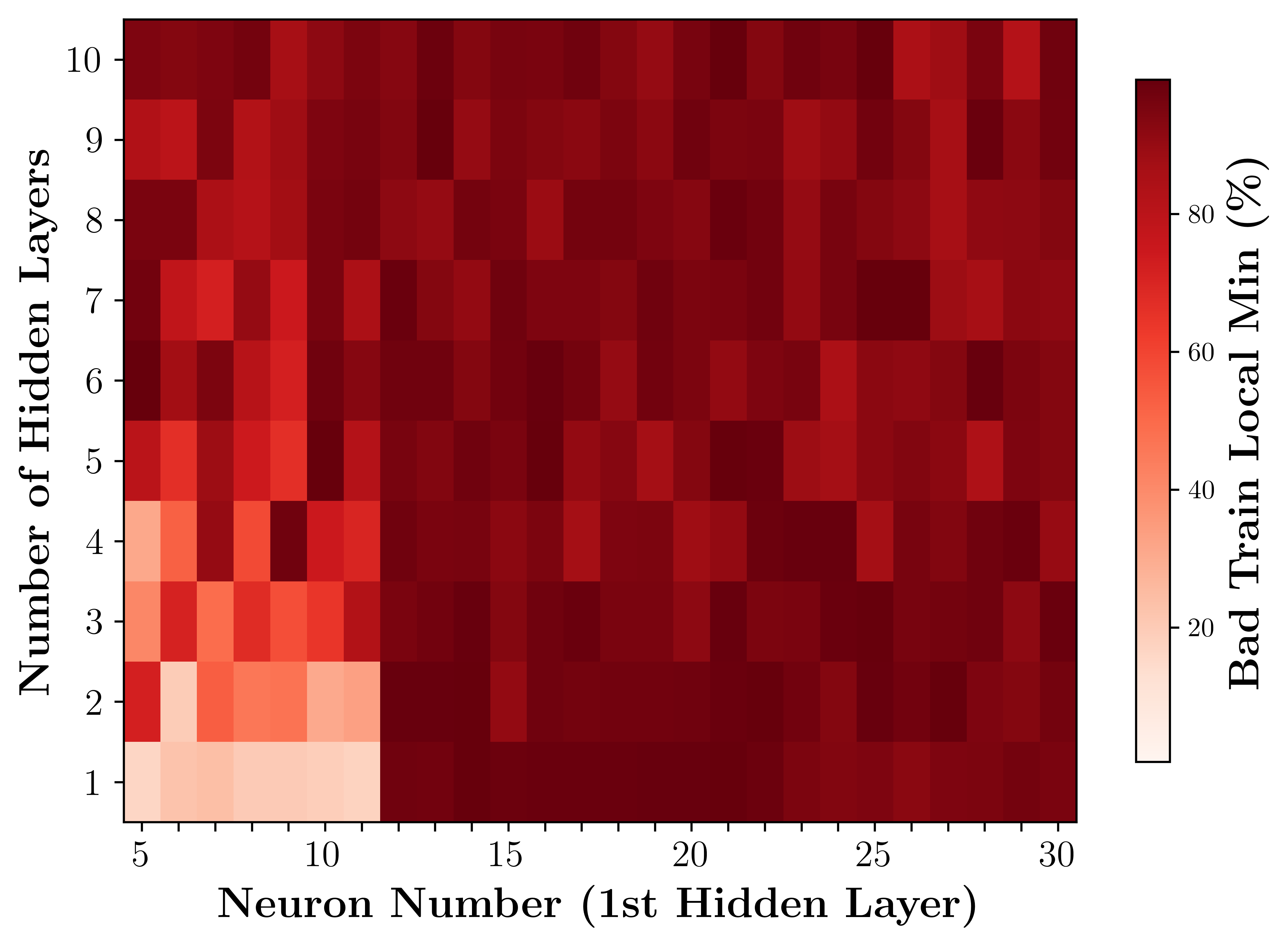}  
    \captionsetup{justification=raggedright,singlelinecheck=false}
    \caption{Bad local minima proportion in SinNN 's training set.}
    \label{fig:20w0_200seeds_0.002lr_300epoch_threshold1_train_heatmap}
\end{figure}

\begin{figure}[!htbp]
    \centering
    \includegraphics[width=0.95\linewidth]{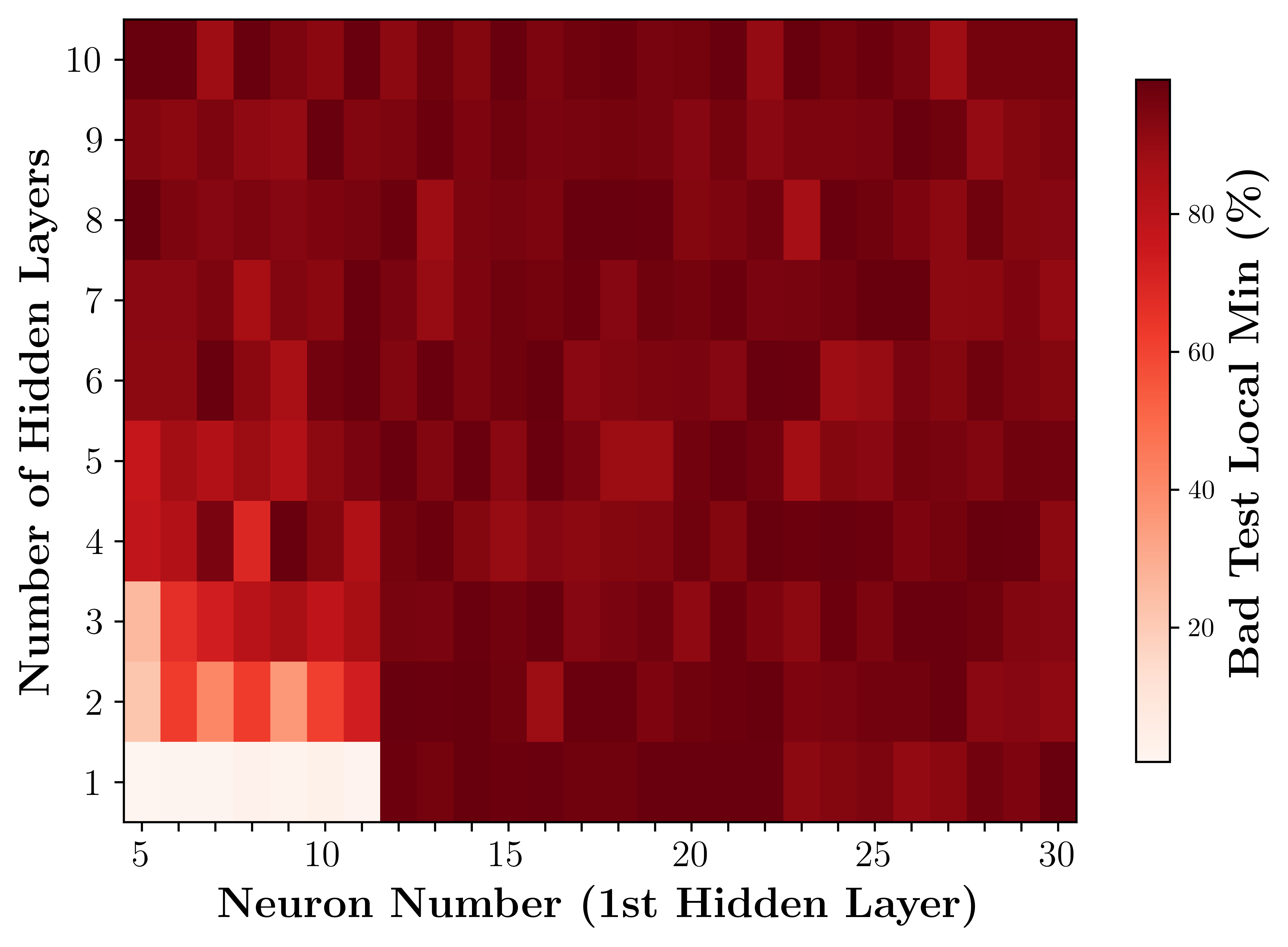}  
    \captionsetup{justification=raggedright,singlelinecheck=false}
    \caption{Bad local minima proportion in SinNN 's test set.}
    \label{fig:20w0_200seeds_0.002lr_300epoch_threshold1_test_heatmap}
\end{figure}

Heatmaps presented in~\autoref{fig:20w0_200seeds_0.002lr_300epoch_threshold1_train_heatmap} and ~\autoref{fig:20w0_200seeds_0.002lr_300epoch_threshold1_test_heatmap}, 
demonstrate that the likelihood of encountering bad local minima is remarkably high across most network configurations, on both training and test sets, indicating that the issue is inherent to the optimisation landscape of the model. In many settings, more than 95\% of the training runs converge to bad local minima, highlighting the severity of this bad local minima problem. However, a distinct region appears in the lower-left corner of the heatmap, where the fraction of bad local minima is much lower. Upon further inspection, we find that the global minimum losses in this region are significantly higher than those in the rest of the plot. This suggests that these configurations suffer from underfitting due to insufficient model capacity (i.e., too few neurons and layers). Consequently, because our definition of a ``bad” local minimum requires the loss to be at least twice the global minimum, some cases that would ordinarily be categorised as bad local minima are not identified.

\begin{figure}[!htbp]
    \centering
    \includegraphics[width=0.95\linewidth]{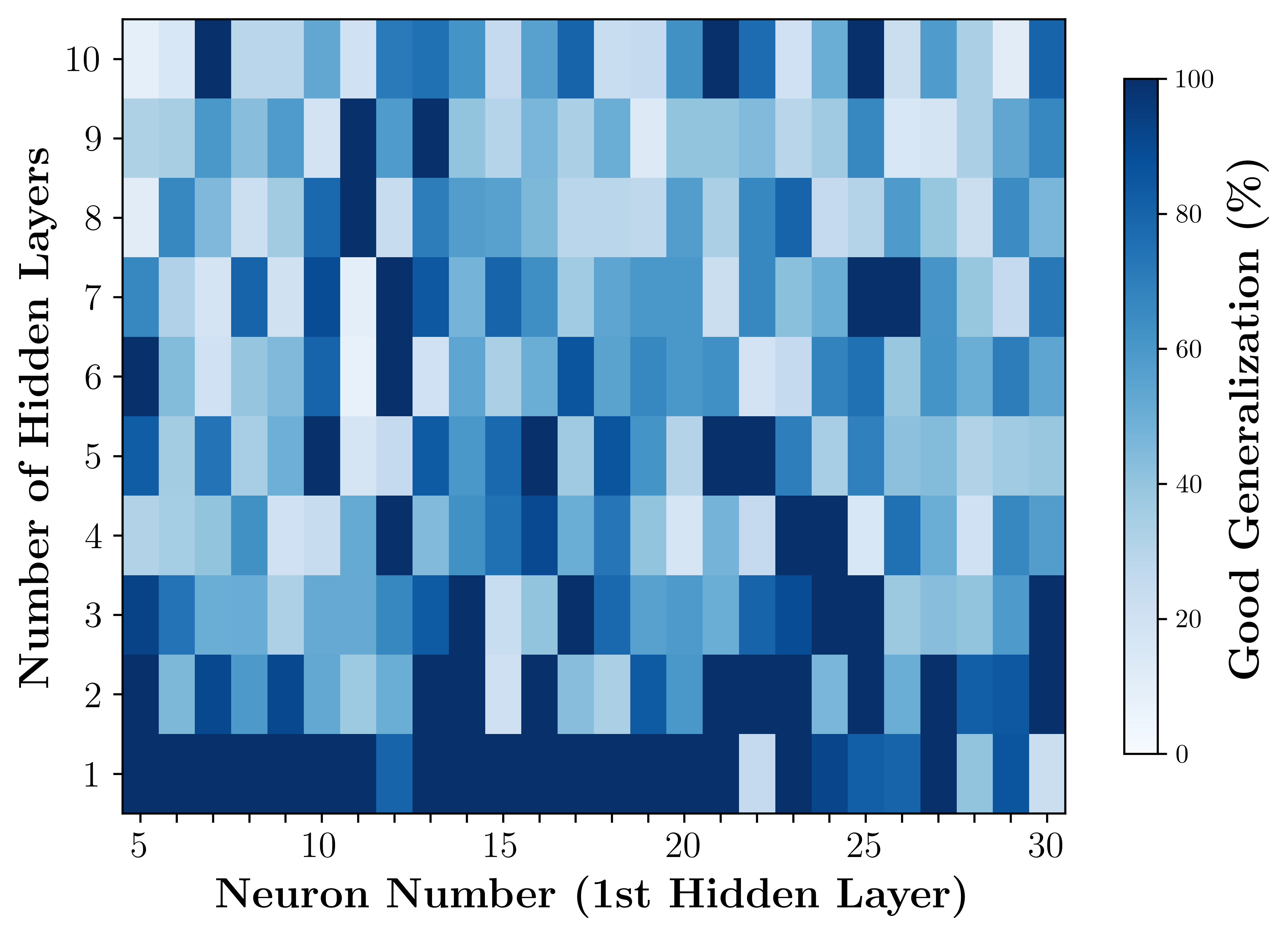}  
    \captionsetup{justification=raggedright,singlelinecheck=false}
    \caption{Good generalisation rate of different SinNN configurations.}
    \label{fig:20w0_200seeds_0.002lr_300epoch_threshold1_good_heatmap}
\end{figure}

As shown in ~\autoref{fig:20w0_200seeds_0.002lr_300epoch_threshold1_good_heatmap}, we also observe that when a model approaches the estimated global minimum on the training set, it is very likely to achieve similarly good performance on the test set. This indicates that trained models with global minima often generalise well, and underscores the importance of successfully reaching the global minimum during training.


\subsection{Bad local minima in large DSinNNs}
To illustrate how the discrete sinusoidal neural network (DSinNN) operates, we consider an example analogous to the continuous case: fitting the target function 
\begin{equation}
    y = \operatorname{round}\left(\sum^5_{k=1}\sin{kx}\right),
\end{equation}
where $\operatorname{round}(\cdot)$ denotes rounding to the nearest integer. The DSinNN adopts the same ``decreasing" architecture, where the number of neurons in each hidden layer decreases progressively after the first. Specifically, we define the number of neurons in the first hidden layer and the total number of hidden layers. We then uniformly sample 200 input-output pairs $(x, y)$ from the interval $[-7\pi, 7\pi]$, with 80\% randomly selected for training and the remaining 20\% used for testing. Notably, we deliberately avoid normalising the input data to $[-1, 1]$. In DSinNN, where direct frequency modulation via weight magnitudes is no longer feasible due to discretisation, we instead leverage a broader input domain to introduce richer periodic variation across the network. While this does not provide the same level of frequency control as in the continuous setting, the increased input range effectively exposes the network to a wider span of phase and periodic patterns. This may help deeper layers to capture higher-frequency behaviour through composition, making the broadened input domain a simple yet potentially impactful strategy under discretisation constraints.

The weight initialisation scheme closely follows that of the continuous SinNN. For the first layer, weights are drawn from $\mathcal{U}(-1, 1)$, discretised, and then scaled by a smaller factor $\omega_0 = 1$, whereas the continuous case uses a larger $\omega_0$ to induce a broader frequency spectrum. For all subsequent hidden layers, weights are sampled from $\mathcal{U}\left(-\sqrt{6/n}, \sqrt{6/n}\right)$, where $n$ is the number of input connections (fan-in), discretised, and then scaled by a constant $\lambda = \min(\frac{\pi}{2}, \frac{\pi}{2^{m-1}})$, where $m=\lfloor\log_2n\rfloor$. This scaling ensures that a meaningful portion of weighted inputs $\mathbf{w}^\top \mathbf{x}$ fall outside the near-linear regime of the sine function (typically $[-\pi, \pi]$), thereby enhancing the network’s ability to exploit non-linear features as suggested by~\cite{SIREN}.

We note that, in general the loss $L = \frac{1}{N} \sum_x (\hat{y}_x - y_x)^2$. The gradient is
\begin{equation}\label{eq:general-gradient}
    \frac{\partial L}{\partial w_i} = \frac{1}{N}  \sum_x 2 \cdot (\hat{y}_x - y_x) \frac{\partial \hat{y}_x}{\partial w_i}.
\end{equation}
When calculating $ \frac{\partial \hat{y}_x}{\partial w_i}$, the discretisation functions are treated as STE identity function $f(x)=x$. For example, in a hidden-layer neuron, the gradient takes the form:
\begin{equation}
    \frac{\partial \hat{y}_x}{\partial w_i} = \cos (\lambda \text{D}(w_i)z) \cdot z.
\end{equation}
where $z$ denotes the sum of fan-in values from discretised values in $\{-1, 0, 1\}$. When $z$ is an even integer and the scaling factor $\lambda = \frac{\pi}{2^{n'}}$, the argument to the cosine function can become a multiple of $\frac{\pi}{2}$, resulting in a zero gradient. For those scenarios, weights are stuck at their initial values, no matter what their losses are. To address this, we add a small constant $0.1$ inside the sine function, so that each neuron's output becomes $\text{D}'(\sin(\lambda \text{D}(w_i)z + 0.1))$, where $\text{D}$ is defined in~\autoref{eq:D} and $\text{D}'$ is given by
\begin{equation}
\text{D}'(a) \equiv
\begin{cases} 
-1 & \text{if } a < -0.1 \\
1 & \text{if } a > 0.1 \\
0 & \text{otherwise}. \\

\end{cases}
\label{eq:D'}
\end{equation}

Because the addition of a small constant does not change the outcome of the discretisation, the loss landscape and optimal solution remain unchanged. At the same time, this approach preserves architectural simplicity and training stability.

The training configuration includes an initial weight scaling factor $w_0 = 1$, 200 random seeds, a learning rate of 0.002, and 300 training epochs. A local minimum is considered ``bad" if its final loss is significantly larger than the global minimum; the criterion for distinguishing ``bad" minima is the same as in the continuous SinNN case.

\begin{figure}[!htbp]
    \centering
    \includegraphics[width=1.0\linewidth]{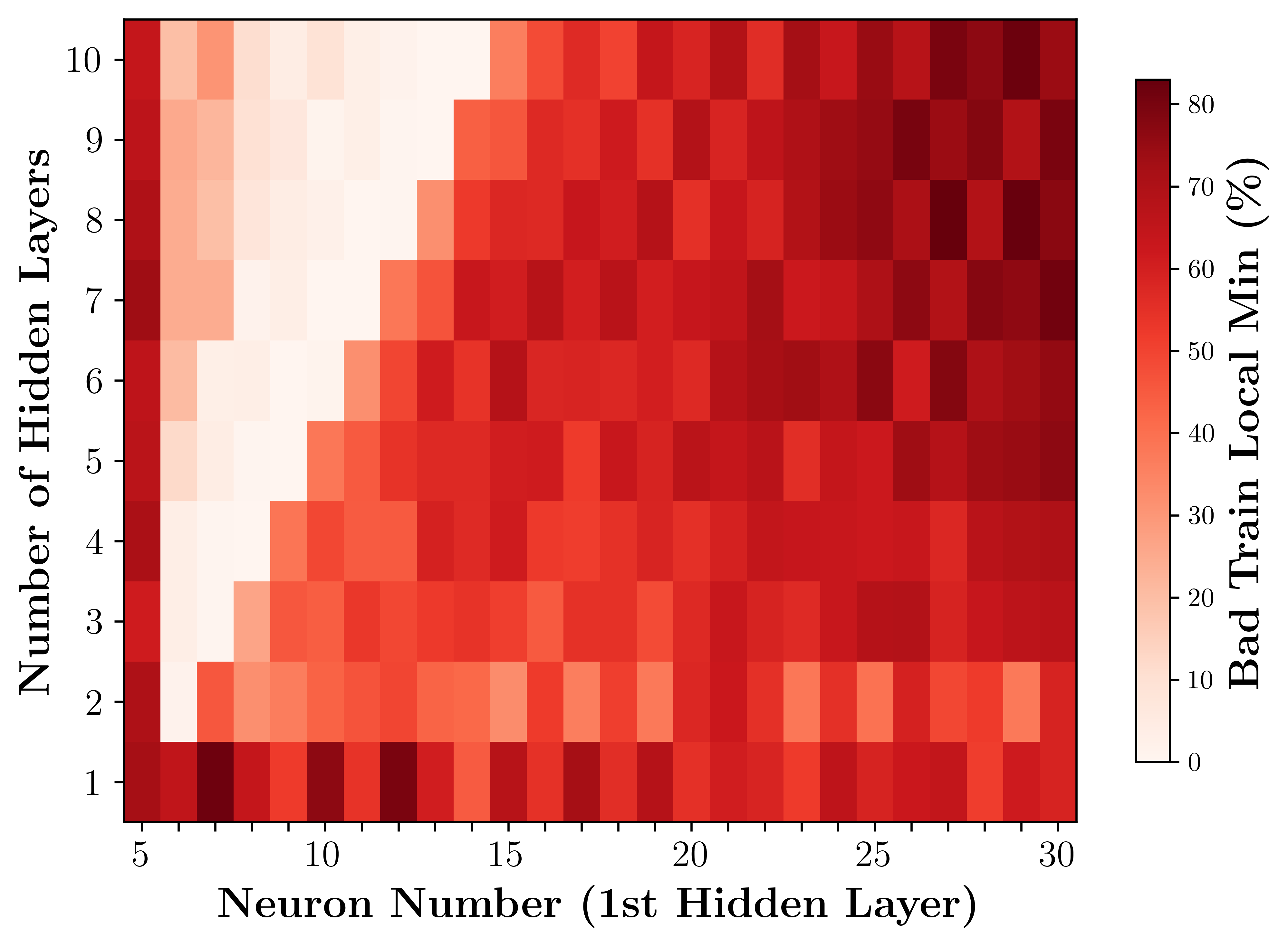}  
    \captionsetup{justification=raggedright,singlelinecheck=false}
    \caption{Bad local minima proportion in DsinNN's training set.}
    \label{fig:200data_Largerange_1w0_200seeds_0.002lr_300epoch_threshold1_train_test_good_train_heatmap}
\end{figure}

\begin{figure}[!htbp]
    \centering
    \includegraphics[width=1.0\linewidth]{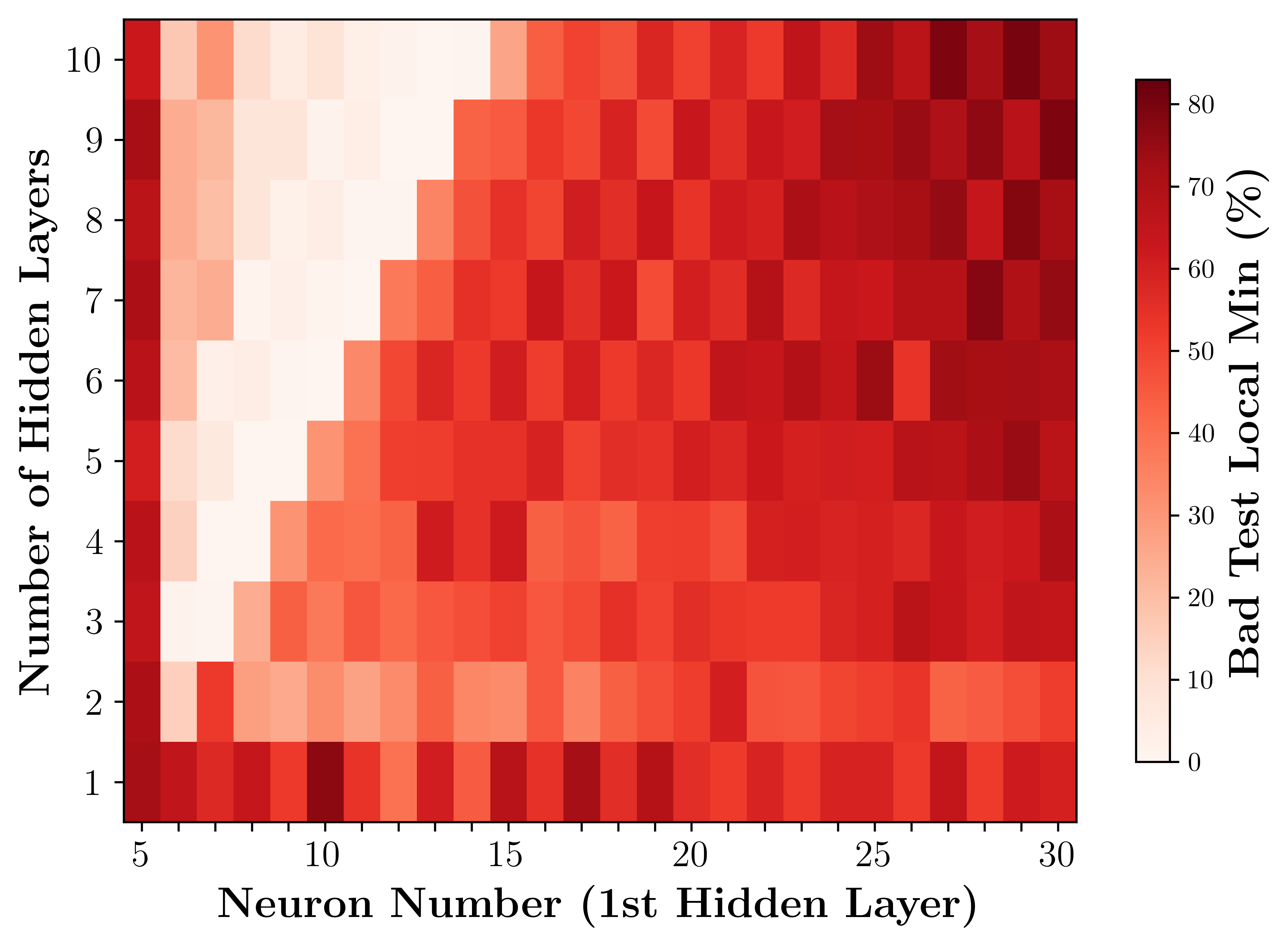}  
    \captionsetup{justification=raggedright,singlelinecheck=false}
    \caption{Bad local minima proportion in DsinNN's test set.}
    \label{fig:200data_Largerange_1w0_200seeds_0.002lr_300epoch_threshold1_train_test_good_test_heatmap}
\end{figure}

We analyse the behaviour of DSinNN across varying network configurations. As shown in our DSinNN results~\autoref{fig:200data_Largerange_1w0_200seeds_0.002lr_300epoch_threshold1_train_test_good_train_heatmap} and~\autoref{fig:200data_Largerange_1w0_200seeds_0.002lr_300epoch_threshold1_train_test_good_test_heatmap}, the proportion of training runs that converge to bad local minima remains high, a trend also observed on the test set. We observe that the upper left regions of the bad local minima heat maps for training and test sets turn ``white." Upon closer inspection, these network configurations do have bad local minima, but their losses are only about 1.5 times the estimated global minimum—below our predefined threshold. As a result, some cases that would otherwise be identified as bad local minima are not detected.

\begin{figure}[!htbp]
    \centering
    \includegraphics[width=1.0\linewidth]{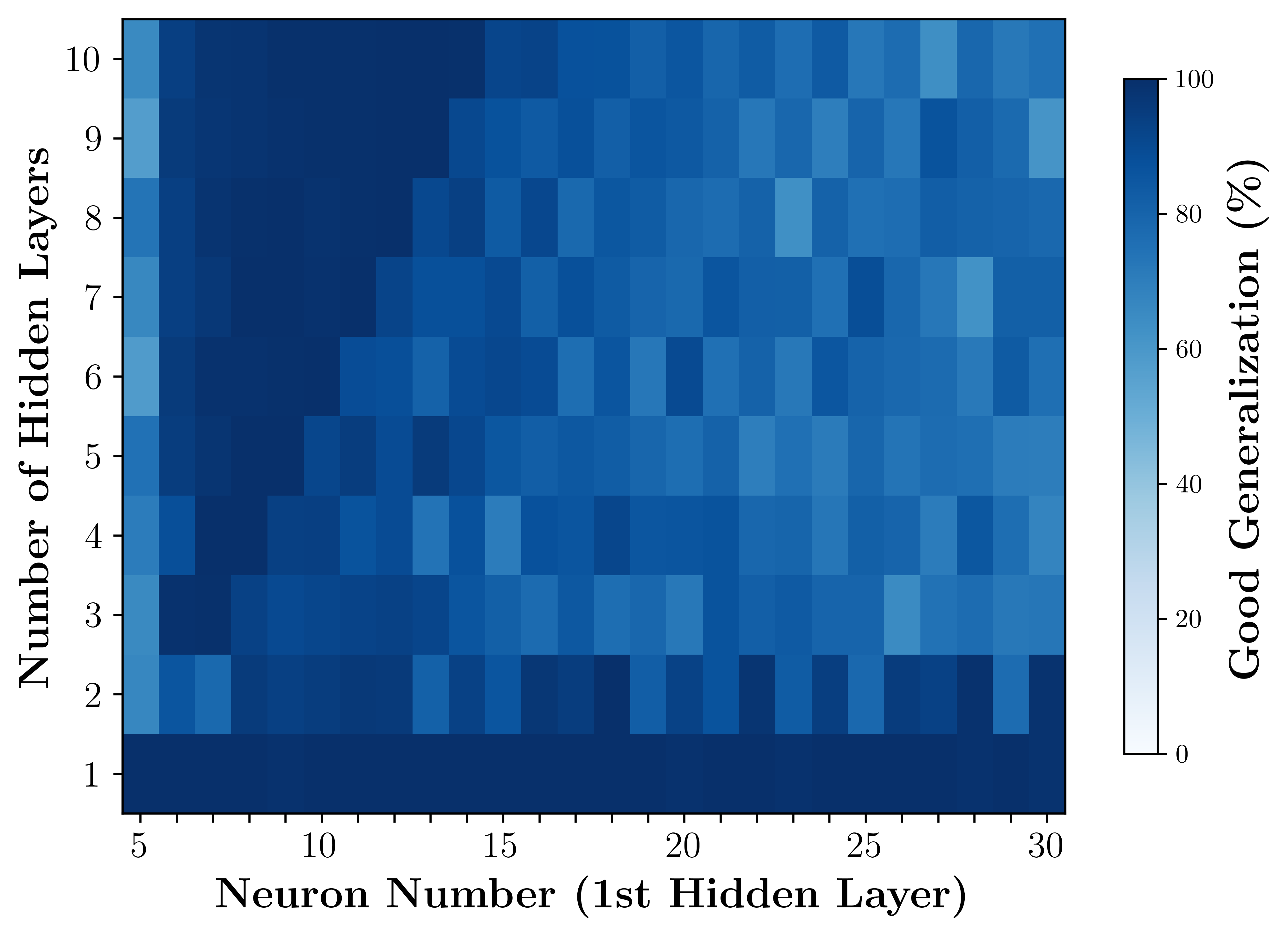}  
    \captionsetup{justification=raggedright,singlelinecheck=false}
    \caption{Proportion of DsinNN's training-set ``good" models that also achieve good performance on DsinNN's test set.}
    \label{fig:200data_Largerange_1w0_200seeds_0.002lr_300epoch_threshold1_train_test_good_good_heatmap}
\end{figure}

Furthermore, as with SinNN, our results in~\autoref{fig:200data_Largerange_1w0_200seeds_0.002lr_300epoch_threshold1_train_test_good_good_heatmap} demonstrate that, for DSinNN, models approaching the estimated global minimum on the training set typically generalise well to the test set. This suggests that achieving the global minimum during training is highly advantageous, as it generally results in strong generalisation.

In addition, our current definition of DSinNN uses 1 bit to represent each weight. If we increase to $n$-bit representations (e.g., using 2 or 3 bits), each weight can take on more possible values, and the network’s parameter space becomes a much finer grid. The same approach can also be applied to activations. As a result, the loss landscape is smoother and more closely approximates the continuous landscape of SinNN.


\section{Quantum Training Algorithm for DSinNN (QSinNN)}
\label{sec:Qtraining}

\begin{figure*}[!htpb]
    \centering
    \adjustbox{trim={3.9cm} {12.5cm} {5.2cm} {2.7cm},clip,scale=1.8}{ 
    \includegraphics[width=1.0\linewidth,page=1]{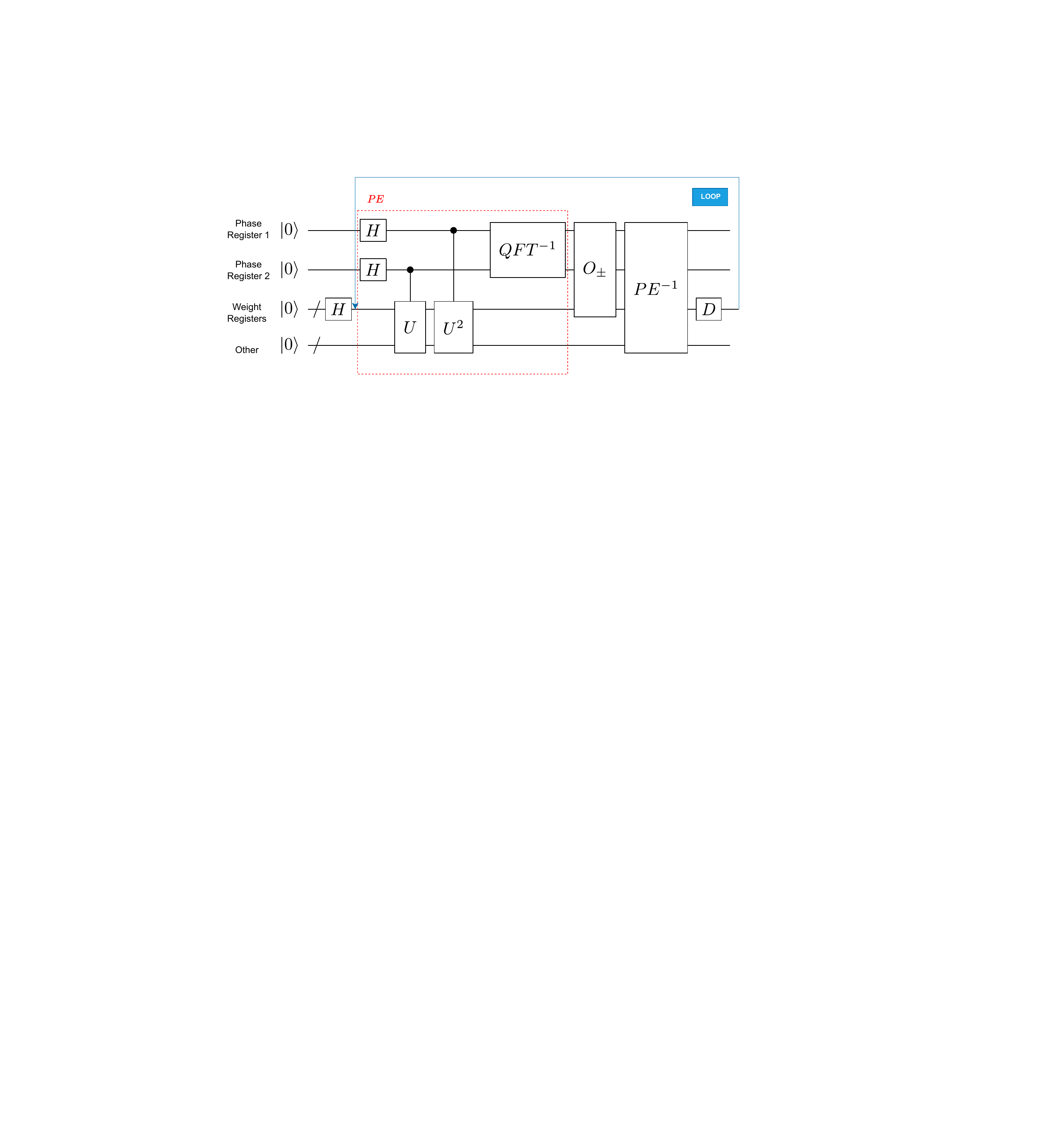}
    }
    \captionsetup{justification=raggedright, singlelinecheck=false}
    \caption{ \textbf{Quantum training for DSinNN (QSinNN).} $PE$ is phase estimation, which we illustrate with two phase registers for simplicity. $O_{\pm}$ is the oracle unitary, which changes amplitudes of states from 1 to 1 or -1 accordingly. $D$ is the diffusion operator which amplifies ``good" weight states.}
    \label{fig:QSinNNGeneral}
\end{figure*}

\begin{figure*}[!htpb]
    \centering
    \adjustbox{trim={2.4cm} {3.4cm} {3.4cm} {3.4cm},clip,scale=1.30}{ 
    \includegraphics[width=1.0\linewidth,page=1]{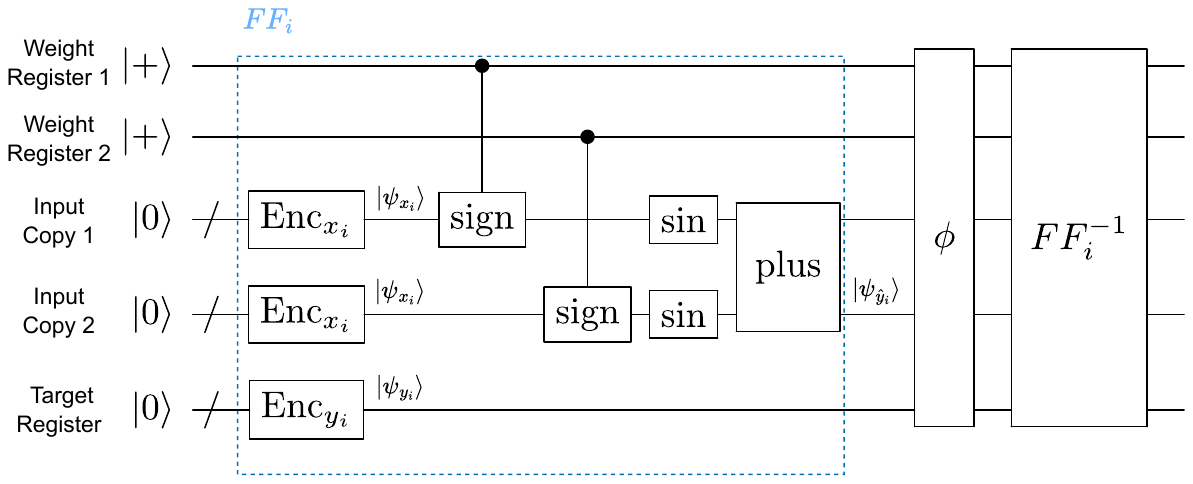}
    }
    \captionsetup{justification=raggedright, singlelinecheck=false}
    \caption{ \textbf{$U_i$ used in unitary $U$ of phase estimation.} $U = U_N \cdots U_2 U_1$, where $N$ is the number of training data pairs. $\text{Enc}_{x_i}, \text{Enc}_{y_i}$ are Pauli $X$ gates to prepare input and target states. The controlled sign gate flips the sign qubit of input qubits if and only if the control qubit is $\ket{1}$. The number of copies of input states is the same as the number of hidden neurons. $\sin,$ plus, and $\phi$ are the quantum $\sin$, plus, and checker gates, whose ancilla qubits are omitted in the drawing. $FF_i$ stands for feed-forward unitary for the $i$-th training data pair.
    }
    \label{fig:QSinNNUnitary}
\end{figure*}

The procedures of the quantum learning algorithm proceed as follows. \autoref{fig:QSinNNGeneral} illustrates the general training procedures, while \autoref{fig:QSinNNUnitary} depicts the unitary gate $U_i$ for processing a single data pair. The overall unitary $U$ (as shown in \autoref{fig:QSinNNGeneral}) is constructed as $U = U_N \ldots U_2 U_1$, where each $U_i$ corresponds to one data pair. For illustration, the example shown is for a network with one hidden layer and two neurons.

\begin{enumerate}
    \item \textbf{Create a uniform superposition of weights.} Each weight register represents one trainable parameter in the classical DSinNN model. The weights are initialized in the state $|0\rangle^{\otimes n}$ and transformed by Hadamard gates into $|+\rangle^{\otimes n}$, where $|+\rangle = \frac{1}{\sqrt{2}}(|0\rangle + |1\rangle)$, where $n$ is the number of weight qubits.

    \item \textbf{Entangle weight registers with phase registers via phase estimation.} The number of phase registers depends on the precision required to distinguish best weight values from other weight values. The unitary operator used in phase estimation simulates the forward passes of classical DSinNN on all training data pairs. Explicitly, $U = U_N \cdots U_2 U_1$, where $N$ is the number of training data pairs. \autoref{fig:QSinNNUnitary} describes each $U_i, i = 1, 2, \dots, N$ in the following steps:

    \begin{enumerate}
        \item $\text{Enc}_{x_i}, \text{Enc}_{y_i}$ encode the $i$-th input and target into corresponding quantum states. Encoding details are described in \hyperref[subsec:quantum-encoding]{Section A}.

        \item Multiplication of weight and input is implemented through a controlled sign gate, i.e. sign qubit of input is flipped if and only if the controlling weight qubit is in state $\ket{1}$.

        \item Quantum plus circuit and quantum sine circuit simulate the classical addition of weighted inputs and the classical sine function. Details are described in \hyperref[subsec:quantum-sine-gate]{Section B} and \hyperref[subsec:quantum-plus-gate]{Section C}. For larger networks (i.e., increased number of neurons and hidden layers), one simply needs to initialize more qubits for the copies of input data and add the corresponding quantum sine circuits and plus circuits according to the neural connectivity between layers. The final output is a quantum state of the DSinNN predicted value. These three steps (a)-(c) constitute a quantum forward pass for $i$-th training data pair, denoted as $FF_i$.

        \item Compare quantum forward pass outputs with target registers through a quantum checker gate $\phi$. For details see \hyperref[subsec:quantum-checker-gate]{Section D}.

        \item Apply inverse of quantum forward pass $FF_i^{-1}$.
    \end{enumerate}

    \item \textbf{Apply the oracle unitary $O_{\pm}$ to invert the amplitudes.} This operation inverts the amplitudes of entangled weight states according to phase registers. Further details are provided in \hyperref[subsec:oracle]{Section E}. 

    \item \textbf{Disentangle weight registers from others by inverse of phase estimation $PE^{-1}$.} This resets all registers to zero states except weight registers. Therefore, weight registers are disentangled from the rest, ending in a pure state.

    \item \textbf{Apply diffusion operator $D$ on weight registers.} 
    \begin{equation}
        D = H^{\otimes n}(2|0^n \rangle\langle 0^n|-I_n)H^{\otimes n},
    \end{equation}
    It increases the amplitudes of ``good" weight states.

    \item \textbf{Steps 2 to 5 are repeated several times.} The final weight states are guaranteed to reach the optimal ones with a high probability~\cite{TIGHTBOUND}.
    
\end{enumerate}

\subsection{Quantum encoding of classical data}
\label{subsec:quantum-encoding}

In this subsection, we describe how to encode classical data into quantum states, i.e. the Enc gates in \autoref{fig:QSinNNUnitary}. We consider cases when every input $x$ can be written in the following form:
\begin{equation}
\label{eq:x=k}
    x = k \cdot \frac{\pi}{2^n}, \hspace{.2cm} k \in \mathbb{Z}, \hspace{.2cm} n \in \mathbb{N}.
\end{equation}
Given a whole dataset, a fixed integer $n$ is chosen as the smallest natural number such that for each input $x$ there exists an integer $k$ satisfying Eq.~\eqref{eq:x=k}. For any integer $k$ except 0, it satisfies:
\begin{equation}
\label{eq:k=b}
    k = (-1)^{\frac{1 - \text{sign}(k)}{2}} \sum_{i=0}^{m - 1} b_i 2^i, \hspace{.2cm} b_i \in \{0, 1\}, \hspace{.2cm} m \in \mathbb{N}.
\end{equation}
Hence any nonzero integer $k$ corresponds to a unique bit string $b_{\text{sign}} b_{m-1} b_{m-2} \cdots b_0$, where $b_{\text{sign}} = \frac{1 - \text{sign}(k)}{2} \in \{0, 1\}$. The number of qubits to encode $|k|$ is $m = \max \{\lceil \text{log}_2 |k|_{\max} \rceil, n\}$, where $|k|_{\max}$ is the largest $|k|$ for all input data. The quantum state $\ket{\psi_x}$ encoding classical input $x$ is just a pure state of $m+1$ qubits:
\begin{equation}
\ket{\psi_x} = \ket{b_{\text{sign}} b_{m - 1} b_{m - 2} \cdots b_0}.
\end{equation}
$\ket{\psi_x}$ can be easily created by applying Pauli X gates to $\ket{0}^{\otimes m + 1}$. When input $x = 0$, its quantum encoding $\ket{\psi_x}$ could be either $\ket{000\cdots 0}$ or $\ket{100\cdots 0}$. 

We give an example to illustrate the quantum encoding of input. Consider input $x = \frac{9 \pi}{8}$. 
\begin{align*}
    n &= \text{log}_2 8 = 3, \\
    m &= \max \{\lceil \log_2 9 \rceil, n\} = 4, \\
    b_{\text{sign}} &= \frac{1 - \text{sign}(9)}{2} = 0, \\
    9 &= 1 \cdot 2^3 + 0 \cdot 2^2 + 0 \cdot 2^1 + 1 \cdot 2^0.
\end{align*}
Finally we get $\ket{\psi_x} = \ket{01001}$.

Since the target $y$ is a discrete integer, its quantum encoding is straightforward using binary encoding: positive integers are represented in standard binary, while negative integers differ only in the leading sign qubit. Such states can also be easily prepared using Pauli $X$ gates.

\subsection{Quantum sine circuit}
\label{subsec:quantum-sine-gate}

\begin{figure}[!htpb]
    \centering
    \adjustbox{trim={2.2cm} {3.2cm} {3.0cm} {3.3cm},clip,scale=1.6}{ 
    \includegraphics[width=1.0\linewidth,page=1]{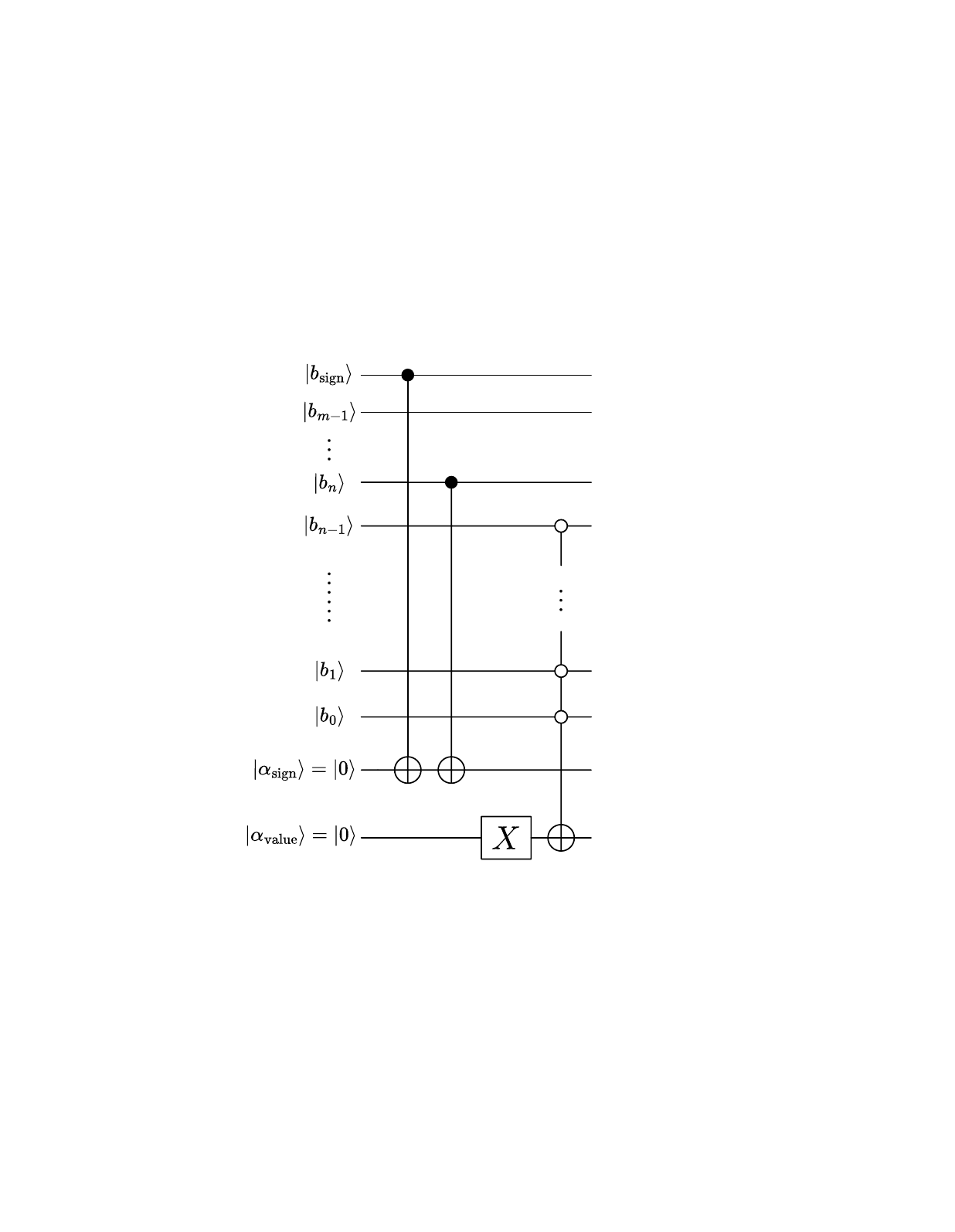}
    }
    \captionsetup{justification=raggedright, singlelinecheck=false}
    \caption{ \textbf{Quantum sine circuit.} It simulates the classical function $\text{D}'(\sin(z))$, assumimg $m>n$. $\ket{\psi_z} = \ket{b_{\text{sign}} b_{m - 1} b_{m - 2} \cdots b_0}$ is the quantum encoding of $z$. $\ket{\alpha_\text{sign}} \otimes \ket{\alpha_\text{value}}$ is the final output representing $\text{D}'(\sin(z))$. $\ket{\psi_z}$ is discarded after this operation.}
    \label{fig:quantum-sine-gate}
\end{figure}

Here we describe the quantum circuit to simulate the classical sine function followed by the discretise function, i.e. D$'(\sin(z))$, where $z$  is the input to the hidden neuron in the classical DSinNN model. Note that this gate is a unitary for a classical nonlinear activation function. Its implementation is illustrated in~\autoref{fig:quantum-sine-gate}. This quantum sine circuit realizes the following mapping of quantum states:
\begin{equation}
    \ket{\psi_z} \otimes \ket{0}^{\otimes 2} \mapsto \ket{\psi_z} \otimes \ket{\alpha_\text{sign}} \otimes \ket{\alpha_\text{value}},
\end{equation}
where
\begin{align}
    \ket{\psi_z} &= \ket{b_{\text{sign}} b_{m - 1} b_{m - 2} \cdots b_0}, \label{eq:z=b}\\
    \ket{\alpha_{\text{sign}}} &= \ket{b_{\text{sign}} \oplus b_n}, \\
    \ket{\alpha_\text{value}} &= \ket{b_0 \vee b_1 \vee \cdots \vee b_{n-1}}.
\end{align}
Here $\ket{\psi_z}$ is the quantum encoding of the classical input to the neuron. It has one qubit representing the sign of $z$ and $m+1$ qubits representing the value of $z$, as described in \hyperref[subsec:quantum-encoding]{Section A}. $\oplus$ is addition module 2 and $\vee$ is the boolean OR operator. An explanation for the quantum sine circuit construction is that: \begin{equation}
\begin{aligned}
\label{eq:z=qr}
    |z| &= \sum_{j=n}^{m-1} b_j 2^j \cdot \frac{\pi}{2^n} + \sum_{i=0}^{n-1} b_i 2^i \cdot \frac{\pi}{2^n} \\
    &= q\pi + r, \quad q \in \mathbb{N}, \quad r \in \mathbb{R}, 
\end{aligned}
\end{equation}
where $r$ represents the portion of $|z|$ that is smaller than $\pi$. Therefore, for D$'(\sin(z))$, if $r\ne0$, then $|\text{D}'(\sin(z))|=1$. Equivalently, if $|b_{n-1}\cdots b_2b_1\rangle\ne|0\cdots00\rangle$, then $|\alpha_{\text{value}}\rangle=|1\rangle$. The bit $b_n$ determines whether $q$ is an odd or even number, and together with $b_{sign}$, it finalizes the sign of $\text{D}'(\sin(z))$, denoted by $|\alpha_\text{sign}\rangle$. $\ket{\psi_z}$ can be discarded after this quantum sine circuit. $\ket{\alpha_\text{sign}} \otimes \ket{\alpha_\text{value}}$ represents $\text{D}'(\sin(z))$ and is kept for following processes.

For neurons in deeper hidden layers, the fan-in weighted sum from previous neurons is represented as a binary integer (the method for computing this sum will be described in~\autoref{subsec:quantum-plus-gate}). As in the classical DSinNN, this integer is multiplied by the scaling factor $\lambda = \frac{\pi}{2^{n'}}$. In the quantum setting, this corresponds to applying a quantum sine circuit to the integer, with the circuit design determined by the chosen scaling factor. More specifically, the fan-in weighted sum integer corresponds to $k$ in~\autoref{eq:x=k}, and $n'$ corresponds to $n$.

\subsection{Quantum plus circuit}
\label{subsec:quantum-plus-gate}
Quantum plus circuit simulates $\sum_i \text{D}(w_{ij})z_i$, where $z_i$ is the output of the $i$-th hidden neurons. In \autoref{fig:QSinNNUnitary}, it is denoted as ``plus" gate. Since $z_i$ is one of $\{-1, 0, 1\}$, its corresponding quantum state only needs two registers: one for its sign and one for its value.
\begin{equation}
    \ket{\psi_{z_i}} = \ket{\alpha_{i s}} \otimes \ket{\alpha_{i v}}.
\end{equation}
The quantum plus circuit consists of four steps:
\begin{enumerate}
    \item Encode number of states $\ket{01}$ $(+1)$, and $\ket{11}$ $(-1)$ in $\{\ket{\psi_{z_i}}, i = 1, 2, \cdots\}$ in registers $\ket{p}$ and $\ket{n}$. Essentially $p$ is the number of positive $ \text{D}(w_{ij})z_i$ and $n$ is the number of negative $ \text{D}(w_{ij})z_i$. This can be implemented using quantum adders~\cite{ADD1,ADD2}.
    \item Convert the negative register $\ket{n}$ to its two's complement form (i.e., bitwise negation plus one), yielding $\ket{-n}$.
    \item Use a quantum adder to sum the positive register $\ket{p}$ and $\ket{-n}$, resulting in $\ket{p + (-n)}$. Provided that the result register is wide enough, a leading $\ket{0}$ indicates a non-negative integer (read as standard binary), while a leading $\ket{1}$ indicates a negative result in two’s complement form.
    \item If the leading qubit is $\ket{0}$, no further action is needed. If the leading qubit is $\ket{1}$, subtract $1$ from the register (excluding the leading qubit) and take the bitwise complement of the remaining bits, while keeping the leading qubit as $\ket{1}$. This outcome is consistent with our quantum encoding of classical data (with scale factor 1).

\end{enumerate}

\begin{figure}[H]
    \centering
    \adjustbox{trim={1.6cm} {2.4cm} {2.8cm} {2.4cm},clip,scale=1.4}{ 
    \includegraphics[width=1.0\linewidth,page=1]{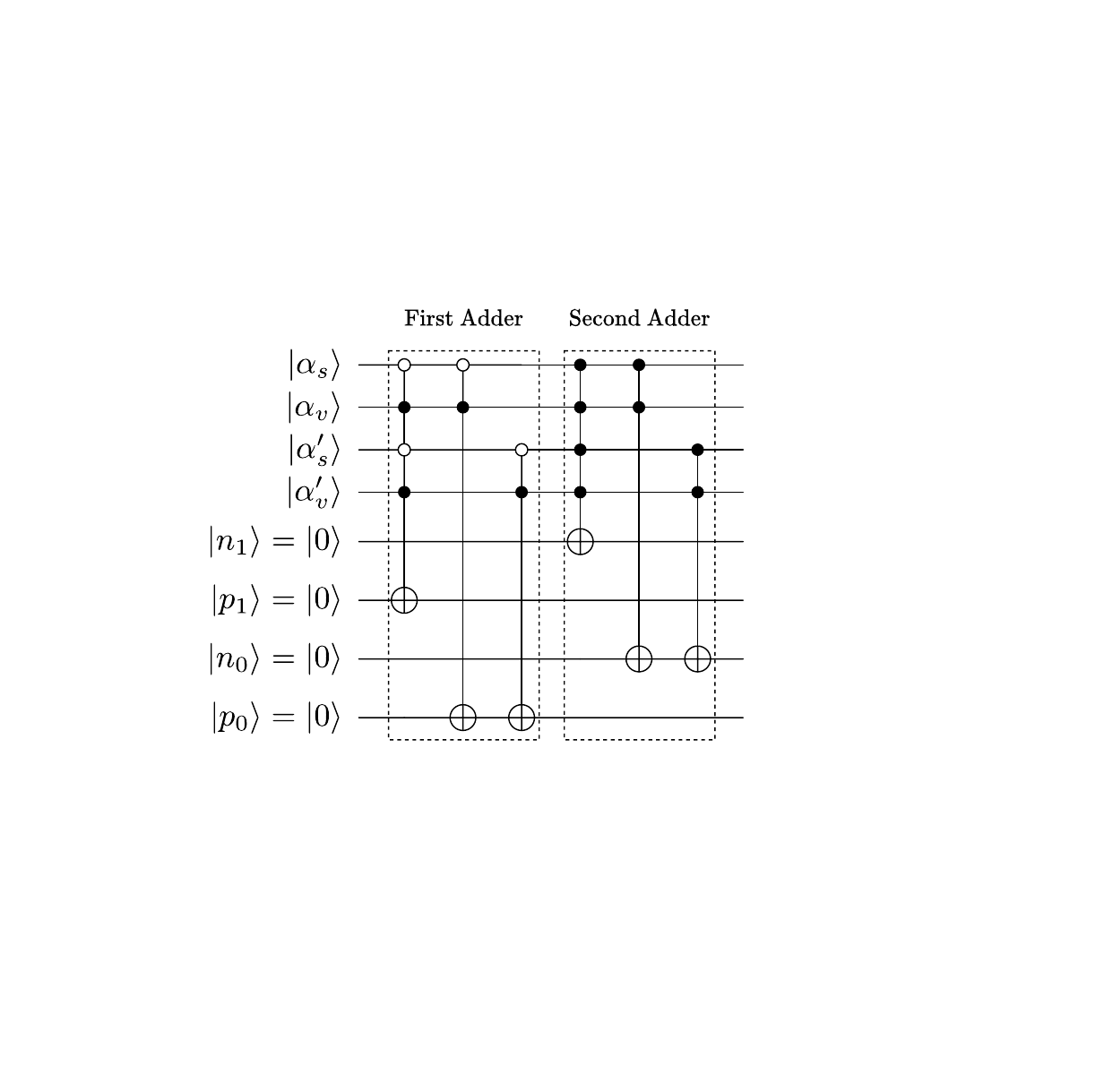}
    }
    \caption{\small \textbf{Two quantum half adders.} Inputs are $\ket{\psi_{z_1}} = \ket{\alpha_{s}} \otimes \ket{\alpha_{v}}$ and $\ket{\psi_{z_2}} = \ket{\alpha'_{s}} \otimes \ket{\alpha'_{v}}$. First adder counts the number of $\ket{01}$ in $\ket{\psi_{z_1}}$ and $\ket{\psi_{z_2}}$ into $\ket{p_1 p_0}$. Second adder counts the number of $\ket{11}$ in $\ket{\psi_{z_1}}$ and $\ket{\psi_{z_2}}$ into $\ket{n_1 n_0}$.}
    \label{fig:HalfAdder}
\end{figure}

\autoref{fig:HalfAdder} illustrates a simple example using two quantum half adders to implement step 1 for $\ket{\psi_{z_1}} = \ket{\alpha_{s}} \otimes \ket{\alpha_{v}}$ and $\ket{\psi_{z_2}} = \ket{\alpha'_{s}} \otimes \ket{\alpha'_{v}}$. The output is $\ket{p_1 p_0} \otimes \ket{n_1 n_0}$. 
\begin{equation}
    \ket{p_1 p_0} = \begin{cases}
        \ket{00} & \text{if } \ket{\psi_{z_1}} \neq \ket{01} \text{ and } \ket{\psi_{z_2}} \neq \ket{01}, \\
        \ket{01} & \text{if } \ket{\psi_{z_1}} = \ket{01} \text{ xor } \ket{\psi_{z_2}} = \ket{01}, \\
        \ket{10} & \text{if } \ket{\psi_{z_1}} = \ket{\psi_{z_2}} = \ket{01}.
    \end{cases}
\end{equation}
\begin{equation}
    \ket{n_1 n_0} = \begin{cases}
        \ket{00} & \text{if } \ket{\psi_{z_1}} \neq \ket{11} \text{ and } \ket{\psi_{z_2}} \neq \ket{11}, \\
        \ket{01} & \text{if } \ket{\psi_{z_1}} = \ket{11} \text{ xor } \ket{\psi_{z_2}} = \ket{11}, \\
        \ket{10} & \text{if } \ket{\psi_{z_1}} = \ket{\psi_{z_2}} = \ket{11}.
    \end{cases}
\end{equation}
If more $\ket{\psi_{z_i}}$ are involved in step 1, additional quantum full adders are needed. These can be constructed easily through combinations of quantum half adders.

\subsection{Quantum checker gate}
\label{subsec:quantum-checker-gate}
\begin{figure}[H]
    \centering
    \adjustbox{trim={1.9cm} {2.6cm} {2.8cm} {2.6cm},clip,scale=1.60}{ 
    \includegraphics[width=1.0\linewidth,page=1]{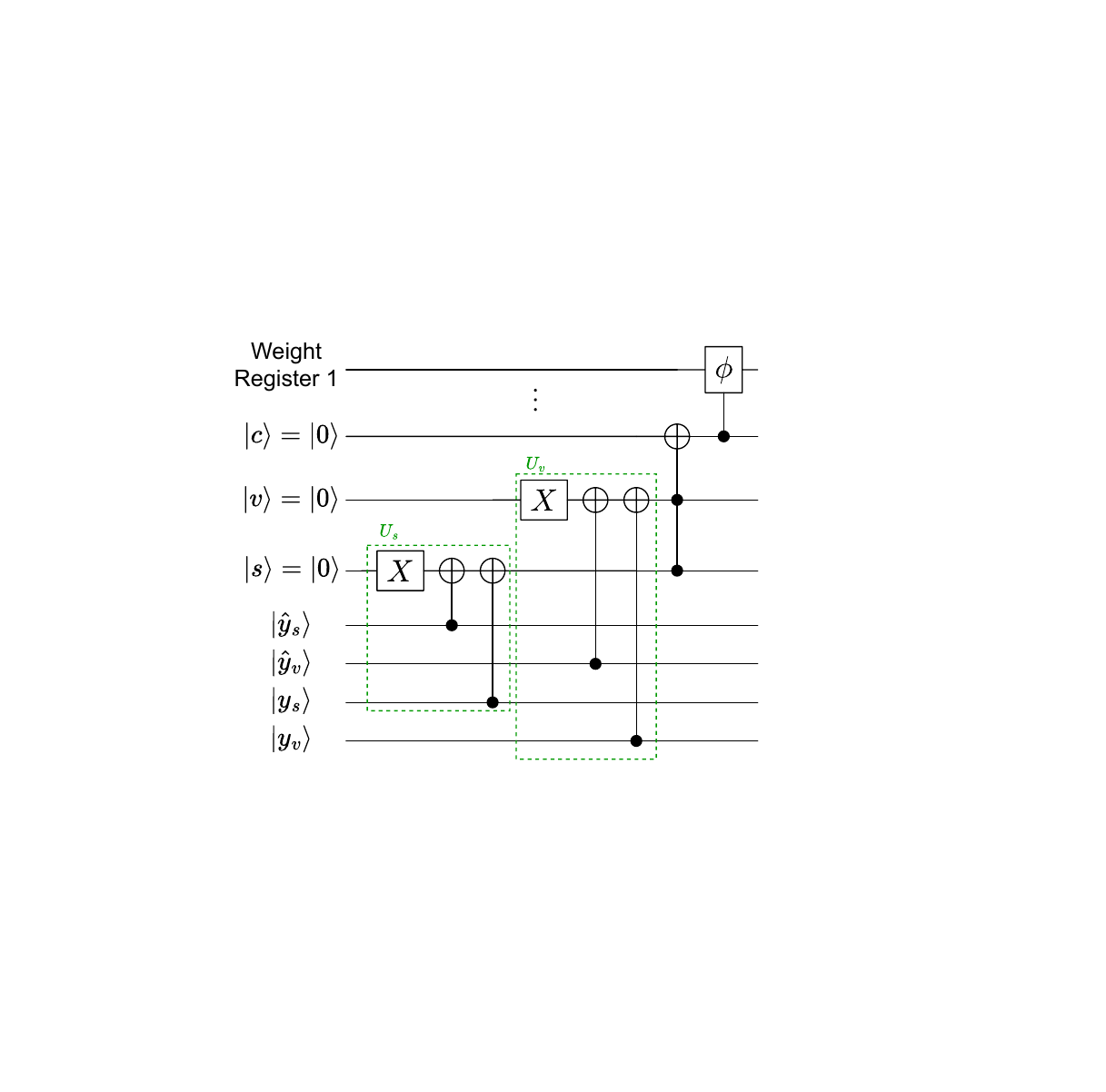}
    }
    \caption{ \textbf{Quantum checker gate.} It checks whether $|\hat{y}_s\rangle=|y_s\rangle$ and $|\hat{y}_v\rangle=|y_v\rangle$. If so, a phase shift $\frac{\pi}{N}$ will be given to weight register 1, where $N$ is the number of training data pairs.}
    \label{fig:checker}
\end{figure}

Quantum checker gate checks whether the model output $\ket{\psi_{\hat{y}}}=\ket{\hat{y}_s \hat{y}_{v_n} \ldots \hat{y}_{v_0}}$ matches the target output $\ket{\psi_y} = \ket{y_s y_{v_n} \ldots y_{v_0}}$ by comparing each corresponding pair of qubits using $U_s$ and $U_{v_i}$. If the outputs match, weight register 1 is rotated by a phase angle $\frac{\pi}{N}$, where $N$ is the number of training data pairs. This phase is denoted as $\phi$ in \autoref{fig:QSinNNUnitary}. For convenience, suppose $\ket{\psi_{\hat{y}}} = \ket{\hat{y}_s \hat{y}_{v}}$ and $\ket{\psi_y} = \ket{y_s y_{v}}$; its implementation is shown in \autoref{fig:checker}. The checker gate does not affect any weight registers other than weight register 1.

If weight states produce correct predictions for all training data, its phase would be $e^{i \pi} = -1$. The phase angle for the weight state is larger and closer to $\pi$ for those whose corresponding model outputs have more correct predictions, compared to those with fewer correct predictions.

\subsection{Oracle gate}
\label{subsec:oracle}

The weights with the phase shift close to \(\pi\) will be considered as the optimal weights. After phase estimation, the information about the phase shift moves to the phase registers. The oracle gate $O_{\pm}$ in \autoref{fig:QSinNNUnitary} shifts phases of optimal weight states by an exact angle $\pi$ according to coupled phase registers. 

More specifically, after phase estimation, phase registers are swapped to an inverse order and expressed as $\ket{j_1 j_2 \cdots j_{N'}}$, representing the estimated state that previously held a phase of $2\pi j = 2\pi (j_1/2+j_2/4+\ldots+j_{N'}/2^{N'})$. If the phase is greater than a predefined threshold $2\pi\tau$:

\begin{equation}
    O_{\pm}|j_1j_2 \cdots j_{N'}\rangle = \begin{cases}
        -\ |j_1j_2 \cdots j_{N'}\rangle,\ \text{if}\  j\ge \tau\\
         \ \ \ |j_1j_2 \cdots j_{N'}\rangle.\ \text{otherwise}\\
    \end{cases}
\end{equation}

\section{Verification and Computational Complexity}
\subsection{Verification}
The overall circuit is organised in a hierarchical manner. The outermost layer is Grover’s algorithm, which consists of two main components: the oracle and the diffusion operator. In our construction, the oracle is realized by a phase estimation routine. One might question whether our circuit truly achieves the intended goal, particularly regarding the design of the unitary gate within the phase estimation loop. To address this concern, we verify the correctness of our design of a unitary gate in a quantum circuit step by step using Dirac notation, as illustrated in~\autoref{fig:QSinNNUnitary}.

Suppose we have $n$ data pairs. We begin by initialising the weight registers in an equal superposition and encoding the first data pair into qubits, along with the necessary ancilla qubits, resulting in the state:
\begin{equation}
\frac{1}{2}\sum_{j,k=0}^{1}|w_j\rangle|w_k\rangle|\psi_{x_1}\rangle|\psi_{x_1}\rangle|0\rangle|\psi_{y_{1}}\rangle.
\end{equation}
Next, the weight states control the sign of the input states, yielding:
\begin{equation}
\frac{1}{2}\sum_{j,k=0}^{1}|w_j\rangle|w_k\rangle|(-1)^{w_j}\psi_{x_1}\rangle|(-1)^{w_k}\psi_{x_1}\rangle|0\rangle|\psi_{y_{1}}\rangle.
\end{equation}
These weighted input states then pass through the quantum sine circuits and are summed, resulting in 
\begin{equation}
\frac{1}{2}\sum_{j,k=0}^{1}|w_j\rangle|w_k\rangle|(-1)^{w_j}\psi_{x_1}\rangle|(-1)^{w_k}\psi_{x_1}\rangle|\psi_{jk,1}\rangle|\psi_{y_{1}}\rangle,
\end{equation}
where $|\psi_{jk,1}\rangle$ encodes the output for the particular combination of $j$ and $k$. Next, $|\psi_{jk,1}\rangle$ is compared with $|\psi_{y_{1}}\rangle$. If the comparison is successful (i.e., the outputs match), the corresponding state acquires a small phase shift of $\frac{\pi}{n}$:
\begin{equation}
\frac{1}{2}\sum_{j,k=0}^{1}e^{i\frac{\pi}{n}\cdot\delta_{j,j^*_1}\cdot\delta_{k,k^*_1}}|w_j\rangle|w_k\rangle|(-1)^{w_j}\psi_{x_1}\rangle|(-1)^{w_k}\psi_{x_1}\rangle|\psi_{jk,1}\rangle|\psi_{y_{1}}\rangle,
\end{equation}
where $j^*_1$ and $k^*_1$ denote the correct indices that yield the target output $|\psi_{y_{1}}\rangle$. Finally, after uncomputation, the state evolves to:
\begin{equation}
\frac{1}{2}\sum_{j,k=0}^{1}e^{i\frac{\pi}{n}\cdot\delta_{j,j^*_1}\cdot\delta_{k,k^*_1}}|w_j\rangle|w_k\rangle|0\rangle|0\rangle|0\rangle|0\rangle.
\end{equation}

This procedure is then repeated for each data pair up to $n$. The optimal $(j, k)$ combination accumulates the largest phase shift, approaching $e^{i\pi}$. Phase estimation is then used to determine whether the phase accumulation of each $(j, k)$ combination surpasses the threshold, which in turn determines whether a ``$-1$" phase shift is ultimately applied to those states.

\subsection{Computational Complexity}
We now analyse the difference in training time between the quantum and classical approaches. 

In a binary neural network, where each weight takes one of two possible values (e.g. -1 or 1), there are $2^M$ possible configurations for $M$ weights. The complexity of a brute-force search to find the optimal binary weights grows exponentially with the number of weights, requiring $N_C^{cl}= N \cdot 2^M$ calls, where $N$ is the number of training data pairs.

There are, under certain assumptions,  guarantees on the rate of convergence of classical gradient descent, though not on reaching a global minimum. It can be shown that $O(N M / \epsilon^2)$ number of iterations achieves convergence to an {\em $\epsilon$-critical point} (including sub-optimal local minima, local maxima, and saddle points) \cite{GD1, GD2}, where $\epsilon$ is the upper bound of the final gradient. That scaling is comparatively fast for fixed $\epsilon$, yet SinNN and DSinNN have many suboptimal minima as demonstrated above (and the cost landscape of DSinNN is not smooth). 
 
For the quantum training method, since the number of ``good" weights is not predetermined, $O(\log (N/\delta))$ times of Grover's search is required, where $\delta$ is the precision needed to find an appropriate threshold $\tau$~\cite{QBNN}. So $O(\log (N/\delta) \sqrt{2^M})$ of Grover's iterations are expected. Each iteration has $O(1)$ phase estimations. The precision of phase estimation is $O(1/N)$, so the number of phase registers is $O(\log N)$ and the number of controlled unitaries is $O(N)$. Each controlled unitary consists of $O(N)$ feed-forward unitaries. In total, the complexity for the quantum training method is $O(N^2 \log (N/\delta) \sqrt{2^M})$~\cite{QBNN}.

The ratio between the classical brute-force search and QSinNN is:
\begin{equation}
\frac{N_C^{cl}}{N_C^{qm}}\approx \frac{2^{M/2}}{N \log (N/\delta)}.
\end{equation}
To make the training process feasible, the ratio of the number of training data pairs $N$ to the number of weights $M$ should be balanced. Since $M$ is typically large, the quantum algorithm often shows significant reductions in training time compared to a global classical search. We expect gradient descent training to be faster than our quantum training algorithm for large numbers of weights; however, it may yield less optimal results, such as getting trapped in local minima as described in former sections, whereas the quantum training is guaranteed to reach the global minimum.

\section{Numerical demonstration of quantum training advantage}

\subsection{Simulation specifics}
To enable a direct comparison with the classical training, the same dataset of input-label pairs $(x_i,y_i)$ used in the gradient training of the DSinNN model is used for the quantum training:
\begin{center}
\begin{tabular}{c c c c}
    $\left(-\frac{3 \pi}{2}, 2\right),$ & $\left(-\frac{\pi}{2},-2\right),$ & $\left(\frac{\pi}{2},2\right),$ & $\left(\frac{3 \pi}{2}, -2\right)$.
\end{tabular}
\end{center}
The input data $x_i$ is encoded into three qubits as state  $|\psi_{x_i}\rangle$, following \autoref{eq:x=k} and \autoref{eq:k=b}. Since one $x_i$ is fed to two neurons, as in DsinNN, the total number of qubits representing the input is $n_{\text{input}}=2\times 3=6$. 

Subsequently, the two copies of $|\psi_{x_i}\rangle$, separately pass through sign gates controlled by weight registers, quantum sine circuits, and then get combined by the quantum plus circuit to output a 3-qubit prediction $|\psi_{\hat{y}_i}\rangle$. 

The quantum plus circuit in~\autoref{subsec:quantum-plus-gate} is implemented using a predefined summation matrix. For example, the sum of two 2-qubit binary numbers is expressed as a 3-qubit binary number, and this matrix maps each possible summation result by the multiplication with a $7\times7$ matrix. The dimension 7 is derived from $2 \times2 +3 = 7$. 

For the comparison between output $|\psi_{\hat{y}_i}\rangle$ and target data $|\psi_{y_i}\rangle$, we use three qubits to encode the target data as in \autoref{subsec:quantum-encoding} and four ancilla qubits to assist comparing the output and target data qubit by qubit. Weight states are assigned a phase shift of $\frac{\pi}{4}$ each time the corresponding model outputs correctly, as previously stated in~\autoref{subsec:quantum-checker-gate}. We use two phase estimation qubits for phase registers, which precisely estimates phase values from the set $\{0, \frac{\pi}{2}, \pi, \frac{3\pi}{2}\}$. 

The total circuit involves 20 qubits as follows: $n_{\text{PE}} = 2$ phase estimation; $n_{\text{weight}} = 2$ for weight states; $n_{\text{input}} = 6$ for input data and its copy (we simplified the implementation of quantum sine circuits so that it does not need ancilla qubits--see code for details); $n_{\text{sum}} = 3$ for summation of two discrete sine results; $n_{\text{target}} = 3$ for target data;  $n_{\text{ancilla}} = 4$ for checking whether the output is consistent with the target. In total, $n_{\text{total}} = n_{\text{PE}} + n_{\text{weight}} + n_{\text{input}} + n_{\text{sum}} + n_{\text{target}} + n_{\text{ancilla}}= 20$.

We use the QuTiP toolbox~\cite{qutip} to build up the quantum circuit for simulation. The corresponding circuit to realise the quantum training described above is shown in the Supplementary Material.

\subsection{Results}
The weight state $|00\rangle$ correctly reproduces all four target data and while the other three weight states reproduce none of the target data. 
Thus $|00\rangle$ should accumulate a phase shift of $\pi$, and the other weight states should accumulate zero phase shift. 

The QuTiP simulation shows that, after phase estimation, the density matrix of the phase registers becomes $\mathrm{diag(0.75, 0.00, 0.25, 0.00)}$. To interpret this result, recall that the superposition states of the two phase registers can be denoted as $|j_1j_2\rangle$, and the phases are represented as $2\pi j$, with $j$= $0.j_1j_2$= $j_1/2+j_2/4$, where $j_i\in\{0, 1\}$. Thus the phase register density matrix shows that one weight gains $(2\pi)\times(1\times\frac{1}{2}+0\times\frac{1}{4})=\pi$ phase shift, and the other three weight states gain zero phase shift. Then we set our correctness threshold (described in~\autoref{subsec:oracle}) to 4, corresponding to phase threshold $4\times \frac{\pi}{4}=\pi$, which in this case acts trivially, retaining the exact $\pi$ phase shift of the correct weight state. 

After uncomputing the circuit above to disentangle the weight state from the other qubits, the weight state density matrix is:
\[
\begin{bmatrix}
 0.25& -0.25 & -0.25 & -0.25\\ 
 -0.25& 0.25 &0.25  &0.25\\ 
 -0.25& 0.25 & 0.25 &0.25 \\ 
 -0.25&0.25& 0.25 & 0.25
\end{bmatrix}.
\]
That density matrix shows that only $|00\rangle$ gains a -1 factor as desired.  

After the subsequent diffusion operator, $H^{\otimes n}(2|0\rangle\langle 0|-I)H^{\otimes n}$, where $n$ is the number of weight qubits, the final density matrix for the weight qubits becomes $\mathrm{diag(1,0,0,0)}$. Therefore, the weight state chosen is indeed $|00\rangle$ as desired. That weight state does not flip the sign of input data, corresponding to classical $w_1=1$ and $w_2=1$. Thus, the quantum training succeeded in finding the global optimum. The star in \autoref{fig:bad_local_minima_SinNN} depicts the successful quantum training outcome and contrasts it with the classical training outcome.

\section{Discussion}

\subsection{Loss landscape seen by quantum search}
We now argue that the quantum search is associated with a sequence of cost function landscapes with global minima only.

The key unitaries in Grover's search can be interpreted as evolution generated by kinetic and potential energy operators respectively~\cite{GROEXP}. More specifically, in a notation similar to Ref.~\cite{GROEXP}, the time evolution under Grover search is
\begin{equation}
\ket{\psi(x,\tau)}=(DR)...(DR)(DR)(DR)\ket{\psi(x,0)},
\end{equation}
where $D$ is the diffusion operator and $R$ the oracle.\\
$D \equiv$ \scalebox{0.85}
{$\begin{bmatrix}
(1 - iN\epsilon) & i\epsilon & i\epsilon & \cdots & i\epsilon \\
i\epsilon & (1 - iN\epsilon) & i\epsilon & \cdots & i\epsilon \\
i\epsilon & i\epsilon & (1 - iN\epsilon) & \cdots & i\epsilon \\
\vdots & \vdots & \vdots & \ddots & \vdots \\
i\epsilon & i\epsilon & i\epsilon & \cdots & (1 - iN\epsilon)
\end{bmatrix},$
}
and $\epsilon$ arises from discretising the time in the Schr\"odinger equation and is infinitesimal, defined by $dx=\sqrt{\frac{dt}{\epsilon}}$. Moreover,\\
$R \equiv$ \scalebox{0.85}{
$\begin{bmatrix}
e^{-iV(x_1)dt} & 0 & 0 & \cdots & 0 \\
0 & e^{-iV(x_2)dt} & 0 & \cdots & 0 \\
0 & 0 & e^{-iV(x_3)dt}& \cdots & 0 \\
\vdots & \vdots & \vdots & \ddots & \vdots \\
0 & 0 & 0 & \cdots & e^{-iV(x_N)dt}
\end{bmatrix}$
}. 

Thus $R=\exp(-i\sum_jV(x_j)\ket{j}\bra{j}dt)$ is associated with the potential energy generator $\sum_jV(x_j)\ket{j}\bra{j}$ with potential energy landscape $V(x_j)$. The generator of $D$ can then be interpreted as a discretised momentum operator $T$ since $DR$ arose from discretising the Schroedinger equation generated by $T+V$. Thus the initial superposition state of the Grover search can be interpreted as a superposition of initial positions, and the time evolution is the rolling down the hill $V(x_j)$. For Grover search $V(x_j)\in \{0,-\pi\}$ with $-\pi$ being the dip in the hill. For example, if the second state is marked and is thus the dip, then $V=\mathrm{diag(0,-\pi,0,0,...)}$. 

\begin{figure}[H]
    \centering
    \includegraphics[width=1\linewidth]{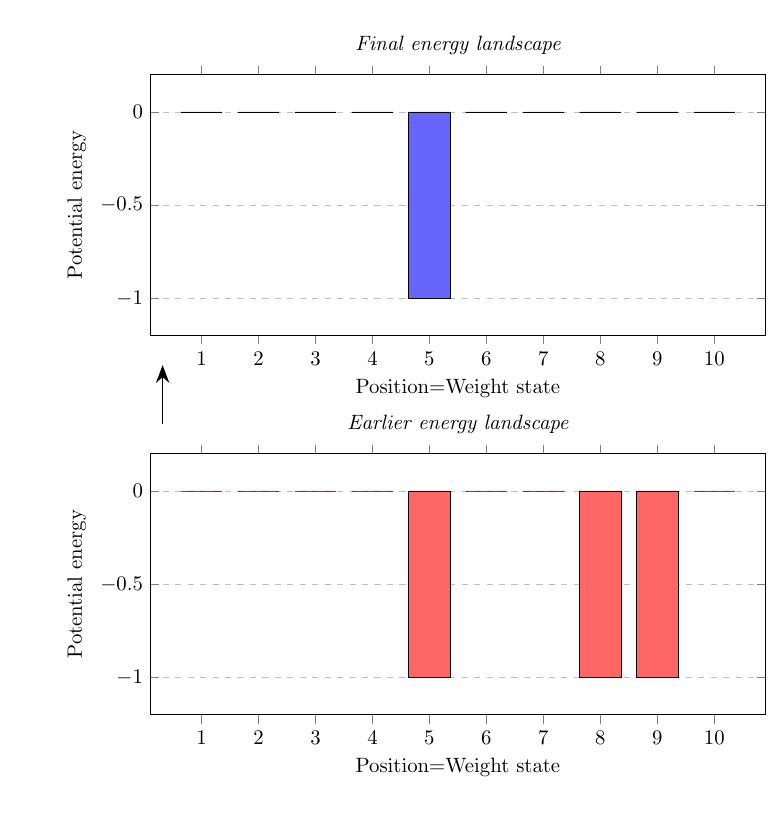}
    \caption{{\bf Energy landscapes associated with quantum search.} Marked elements correspond to global minima of potential energy. Each threshold for which elements are marked is associated with a different landscape, with successive landscapes having fewer and fewer global minima.}
    \label{fig:barlandscapes}
\end{figure}

There is a sequence of potential energy landscapes associated with our NN training approach. In our NN training approach the threshold for which weight strings are awarded a $\pi$ phase is gradually increased. Each threshold is associated with a given landscape, as depicted in Fig.\ref{fig:barlandscapes}. These landscapes are, in line with the above, binary with any minima being global.  

The physics corresponding to quantum search descent has two significant differences from the physics corresponding to classical gradient descent. Firstly, the quantum search starts in a quantum superposition of different positions, whereas classical gradient descent starts in a single location or probabilistic combination of locations. This quantum parallelism enables the system to converge to the state with the lowest `potential energy' in $O(\sqrt{N})$ time, where $N$ is the number of positions. Secondly, the quantum search evolution has, if implemented faithfully physically, no energy dissipation, as it is generated by a time-independent Hamiltonian. In contrast, classical gradient descent can be interpreted as a dissipation of energy to a zero-temperature heat bath (or finite temperature heat bath in the case of simulated annealing) modelled via Monte Carlo evolution. This difference is one way of understanding why the quantum search must be stopped at the right time, when the system is passing through the minimum, whereas classical gradient descent (at zero temperature) has a stable final point.

\section{Summary and Outlook}
In this work we introduce a quantum analogue of sinusoidal neural networks by engineering a quantum sine circuit that realises a discretised sine activation, and we build a quantum optimisation algorithm around it that merges quantum search with phase estimation to guarantee convergence to the global minimum on the training set. We analyse the algorithm’s complexity, verify its operation on a representative example, and, in parallel, design a classical discrete sinusoidal network to establish a fair baseline. Our empirical results show that conventional gradient-descent training for classical sinusoidal nets often settles in poor local minima with high test-set error, whereas our quantum approach avoids these traps, thereby underscoring its potential for more reliable and accurate learning

A key area for development is to minimise the required hardware resources, including the number of CNOT gates. Error correction and mitigation should similarly be investigated. 


\begin{acknowledgments}
We thank Kaiming Bian, Feiyang Liu, Fei Meng, Maria Violaris, Ge Zhang, Jayne Thompson and Mile Gu for discussions. We acknowledge support from the City University of Hong Kong (Project No. 9610623). 
\end{acknowledgments}

\section*{Code availability}
The codes implementing the quantum and classical trainings are available at \url{github.com/zw2788/Quantum-Sine-NN/}.

\section*{Author contributions}
ZW and JH undertook coding and simulations. ZW, JH and OD all contributed to the ideas and in writing the manuscript. 

\bibliography{ref}

\end{document}